\documentclass[journal]{IEEEtranTCOM}
\IEEEoverridecommandlockouts
\usepackage{cite}

\usepackage{graphicx}
\usepackage{subfig}
\usepackage{boxedminipage}
\usepackage{comment}
\usepackage{url}
\usepackage{bm}

\usepackage[cmex10]{amsmath}
\usepackage{amsmath}

\usepackage{url}

\usepackage{times,color}

\begin{document}
%
\title{Optimizing Transmission Lengths for Limited Feedback  with Non-Binary LDPC Examples\thanks{This material is based upon work supported by the National Science Foundation under Grant Numbers 1162501 and 1161822.
 Any opinions, findings, and conclusions or recommendations expressed in this material are those of the author(s) and do not necessarily reflect the views of the National Science Foundation.  
This research was carried out in part at the Jet Propulsion Laboratory,
California Institute of Technology, under a contract with NASA, and JPL Task Plan 82-17473.}}

\author{\IEEEauthorblockN{Kasra Vakilinia, Sudarsan V. S. Ranganathan, Dariush Divsalar*, and Richard D. Wesel}\\
\IEEEauthorblockA{Department of Electrical Engineering,
University of California, Los Angeles, Los Angeles, California 90095\\
*Jet Propulsion Laboratory, California Institute of Technology,
Pasadena, California 91109
}
}

\maketitle

%
%
\begin{abstract}
This paper presents a general approach for optimizing the number of symbols in increments  (packets of incremental redundancy) in a  feedback communication system with a limited number of increments.  This approach is based on a tight normal approximation on  the rate for successful decoding.   Applying this approach to a variety of feedback systems using non-binary (NB) low-density parity-check (LDPC) codes shows that greater than 90\% of capacity can be achieved with average blocklengths fewer than 500 transmitted bits.    One result is that the performance with ten increments closely approaches the performance with an infinite number of increments.  The paper focuses on binary-input additive-white Gaussian noise (BI-AWGN) channels but also demonstrates that the normal approximation works well on examples of fading channels as well as high-SNR AWGN channels that require larger QAM constellations.  The paper explores both variable-length feedback codes with termination (VLFT) and the more practical variable length feedback (VLF) codes without termination that require no assumption of noiseless transmitter confirmation.  For VLF we consider both a two-phase scheme and CRC-based scheme.
 \end{abstract}


\IEEEpeerreviewmaketitle

%
%
\section{Introduction}
\label{sec:Introduction}

The classical results from \cite{Shannon_IT_1956} show that feedback does not increase the capacity of discrete memoryless channels.  However, Polyanskiy et al.\cite{Polyanskiy_IT_2011_NonAsym} and Chen et al. \cite{Chen_2011_ICC} show that capacity can be approached in a smaller number of channel uses using feedback. Polyanskiy et al. \cite{Polyanskiy_IT_2011_NonAsym} introduce random-coding lower bounds for variable-length feedback coding with termination (VLFT) and without termination (VLF), which approach capacity with average blocklengths of hundreds of bits.  A communication system without feedback, on the other hand, requires thousands of bits to closely approach capacity \cite{Polyanskiy_CCR_2010}.  This paper demonstrates practical systems using non-binary low-density parity-check (NB-LDPC) codes that match or exceed the lower bounds of \cite{Polyanskiy_IT_2011_NonAsym}.  Most of the analysis in this paper is not exclusive for NB-LDPC codes, but NB-LDPC codes are used for demonstration because they perform well in the short-blocklength regime (150 to 600 bits) that is of interest.  

In VLFT analysis of \cite{Polyanskiy_IT_2011_NonAsym}, the receiver provides full noiseless feedback to the transmitter. The transmitter sends additional incremental bits until it knows the receiver has decoded the message correctly, resulting in zero probability of error. The ``T" in VLFT stands for termination and corresponds to a noiseless transmitter confirmation (NTC) bit that the transmitter uses to terminate the transmission.  The NTC is transmitted through a channel different from the main communication channel. In contrast, VLF (without the ``T") does not have the advantage of an NTC.  All VLF forward transmissions go over the same noisy channel.  Thus, there is always a nonzero probability of undetected error in VLF.  

VLF and VLFT are examples of hybrid automatic repeat request (HARQ) schemes.
Prior to Polyanskiy et al. \cite{Polyanskiy_IT_2011_NonAsym} and Chen et al.  \cite{Chen_2011_ICC}, HARQ feedback schemes had been studied in great detail in many papers including for example \cite{Shu_Lin_Yu1,Costello_HARQ_IT_1998,Visotsky_RBIR_TCOM_2005, Lott_Soljanin_ITW_2007,Fricke_Reliability_HARQ_TCOM_2009,EminaAllerton2014}.
These papers provide an overview of HARQ, discuss how error correcting codes can be combined with ARQ and demonstrate applications of HARQ.   In particular, \cite{EminaAllerton2014} shows that hybrid ARQ is especially useful in point-to-point scenarios. The coding schemes that are most commonly explored in HARQ systems \cite{Roongta_HARQ_ICC_2003,Roongta_HARQ_WCNC_2004,BCJR_TransIT_1974} are based on convolutional codes (CCs) or a concatenation of turbo and block parity-check codes, where the Bahl-Cocke-Jelinek-Raviv (BCJR) algorithm is used to determine which bit is unreliable and needs to be transmitted in the subsequent transmissions. These works use a genie (equivalent to NTC in VLFT) to terminate transmissions. 

In order to remove the genie and realize a more practical system (equivalent to VLF) \cite{Raghavan_ROVA_TransIT_1998,Fricke_HARQ_NAW_2006,Fricke_Approx_ROVA_VTC_2007,Fricke_Reliability_HARQ_TCOM_2009, Visotsky_RBIR_TCOM_2005,Adam_TB-ROVA} consider reliability-based HARQ using convolutional codes where the transmission terminates when the probability of having a correctly decoded message is high enough. For example, in \cite{Fricke_Reliability_HARQ_TCOM_2009} the reliability metric is based on the average magnitude of the log-likelihood ratios of the source symbols. 

In \cite{Soljanin_LDPC_HARQ_2005,Soljanin_LDPC_HARQ_ITW_2006}, Soljanin et al. study  VLFT HARQ using rate-compatible binary LDPC codes. They use maximum likelihood (ML) decoding analysis to determine the size of incremental transmissions in case of decoding failure. In \cite{Soljanin_LDPC_ITW_2009,Soljanin_LDPC_IT_2012} Soljanin et al. extend their analysis to time-varying binary erasure channels.

Some other high-throughput ARQ schemes use rateless spinal codes as in \cite{Perry_Spinal_Hotnets_2011,Perry_Spinal_Sigcomm_2012}, where hash functions are used for the subsequent coded symbols. In \cite{Romero_MS_Thesis_2014}, Romero uses cyclic redundancy check (CRC) codes to study the performance of spinal codes in VLF setting. Use of polar codes with HARQ is also studied in \cite{Chen_Polar_HARQ_TCOM_2013, Chen_Polar_HARQ_WCNC_2014}. These works present polar-code-based HARQ schemes over binary-input additive white Gaussian noise (BI-AWGN) and Rayleigh fading channels using Chase combining.

The closest work to the analysis presented here is by Pfletschinger et al. in \cite{Pfletschinger_LDPC_HARQ_TWC_2014} which uses rate-adaptive, non-binary LDPC codes in a HARQ scheme over Rayleigh fading channel in the VLFT setting. They present two algorithms that use channel statistics and mutual information to optimize the blocklengths for each transmission to maximize the throughput. Based on channel state information at transmitter, the code rates, modulations, and  maximum number of retransmissions are all optimized prior to initial transmission.  

Chen et al.~\cite{Chen_2011_ICC, Chen_Feedback_Journal_2013} and Williamson et al. \cite{Williamson_ISIT_2012} analyzed a VLFT
scheme based on rate-compatible sphere-packing with an ML
decoder (RCSP-ML) and simulated a VLFT scheme using convolutional codes.
The approximation based on RCSP-ML extends sphere-packing analysis from a single fixed-length code to a family of rate-compatible codes, where each code in the family achieves perfect packing and is decoded by an
ML decoder.
For the 2-dB BI-AWGN channel
with feedback, the convolutional codes achieve about 95\% of the idealized RCSP-ML throughput ($R_{RCSP}$) for average blocklengths up to 50 bits. In \cite{Adam_Tcom_New}, Williamson et al. also analyzed VLF systems for similar blocklengths of up to 100 bits.

However, for average blocklengths of 100 bits and larger,
the throughput of the convolutional code decreases because the frame-error rate performance of the convolutional code degrades as the length of code increases.
As Chen et al. mention in~\cite{Chen_Feedback_Journal_2013}, coding schemes with throughput performance close to
RCSP-ML in VLFT still remain to be identified
for expected latencies (average blocklengths) of 200 to 600 bits. This blocklength regime is important because it is still short enough that feedback provides a real advantage but also long enough that the system can be practical. 

 The primary purposes of this paper are to show how to optimize the lengths of  incremental transmissions and to demonstrate that
NB-LDPC codes with optimized incremental transmissions
can achieve throughputs close to theoretical limits for expected latencies of 150 to 500 bits in the VLFT and VLF settings.  Most of the following analysis is applicable to any coding scheme, but we use NB-LDPC codes to demonstrate the possible performance motivated by \cite{6404676}, which shows that NB-LDPC codes without feedback, perform well in this short-blocklength regime.

In our precursor conference papers \cite{Kasraisit2014,Kasraitw2014} we preliminarily analyzed the performance of NB-LDPC codes in VLFT for a BI-AWGN channel with an SNR of 2 dB with an unlimited number of transmissions  and with the number of transmissions $m$ fixed to be five.  We also considered two-phase VLF system with $m=5$. In VLFT, the non-binary LDPC codes  of \cite{Kasraisit2014} attain 91\% to 93\% of the predicted RCSP-ML throughput for average blocklengths
of 150 to 450 bits.  In a VLF scheme of \cite{Kasraitw2014} incorporating a confirmation phase after each communication phase (hence called ``two-phase"), 92\% of capacity is achieved in less than 500 bits with a maximum of five transmissions.

In this paper, we extend the results of the previous papers to consider a broader range of $m$, the number of possible transmissions.  We also introduce a new VLF system that uses a stopping criterion that incorporates a cyclic redundancy check (CRC). This new system achieves better throughput performance than the schemes of \cite{Kasraisit2014,Kasraitw2014}  for the example BI-AWGN channel with an SNR of 2 dB in the blocklength regime of 150 to 600 bits. For this channel, the CRC-based VLF scheme achieves about 94\% of the capacity with an unlimited number $m$ of transmissions and about 92\% of the capacity with $m=10$.

We also extend these results to a higher-SNR (8 dB) channel and use a larger 16 quadrature amplitude modulation (QAM) constellation. The capacity of the 8 dB 16-QAM AWGN channel is 2.68 bits per symbol. The VLF-with-CRC system with an unlimited number of transmissions achieves a throughput of 2.37 bits per symbol with a frame error probability of less than $10^{-3}$. This throughput corresponds to 88\% of the capacity in the blocklength regime of about 40 16-QAM symbols. Furthermore, we extend the results to a SNR-5dB BI-AWGN fading channel with the channel state information (CSI) available at the receiver. The capacity of this channel is 0.67 bits. The VLF-with-CRC system with an unlimited number of transmissions achieves a throughput of 0.60 with a frame error probability of less than $10^{-3}$. This throughput corresponds to 90\% of the capacity in the blocklength regime of about 140 bits.   

%
%

The rest of the paper proceeds as follows: Section~\ref{sec:VLFTNBLDPC} provides an overview of the VLFT system with NB-LDPC codes and the reciprocal-Gaussian approximation for the probability mass function of the cumulative blocklengths. 
 Section~\ref{sec:VLFT with limited number of transmissions}
presents the sequential differential optimization algorithm (SDO) for optimizing the size of each incremental transmission in VLFT. Section~\ref{sec:VLFCRC} presents a VLF system with CRC and analyzes this system with an unlimited number of transmissions. Section~\ref{sec:VLFCRC_limited} extends the results of Section~\ref{sec:VLFCRC} to the  system with a limited number of transmissions.  Section \ref{sec:VLFoptimization} gives an overview of the two-phase VLF scheme and uses SDO to optimize the cumulative blocklength at each decoding attempt. Section \ref{sec:results} compares the throughput and the expected latency of NB-LDPC and convolutional codes in VLFT and VLF settings. Section \ref{sec:conclusion} concludes the paper.

\section{VLFT with Non-Binary LDPC Codes}
\label{sec:VLFTNBLDPC}

Feedback can facilitate capacity-approaching performance at significantly
shorter average blocklengths than systems without feedback. This improvement is made possible by capitalizing on favorable
noise realizations to decode early. In case of a bad channel realization, the communication rate is lowered by transmitting additional information until the attempted rate matches the instantaneous rate the channel supports.

 In this paper, building on our precursor conference papers \cite{Kasraisit2014,Kasraitw2014}, we use high-rate protograph-based NB-LDPC codes for the initial transmission. See \cite{6404676} for a discussion of protograph-based LDPC
design. These short-blocklength codes are irregular, having mostly degree-2 and a few degree-1 variable nodes. Refer to \cite{Kasraisit2014} for more discussion on the specification of the codes. 

For most of the analysis, the operating SNR in this paper is 2 dB, similar to the work of \cite{Chen_Feedback_Journal_2013,Kasraisit2014,Kasraitw2014}.  However, to emphasize the generality of the approach in this paper, Section \ref{sec:Generality} shows results for higher-SNR AWGN and fading channels.

It is necessary that the initial transmission
has a rate higher than the capacity to take advantage of good channel realizations.
The coding rate is lowered until decoding is successful.  For
example for SNR-2dB BI-AWGN channel, the initial code can have
a rate of 0.75 to 0.8 while the capacity of the channel is 0.685.

We will consider feedback systems that transmit incremental redundancy one bit at a time and also systems that transmit incremental redundancy in multiple-bit increments.  For systems that use multiple-bit increments,  a practical system may limit the maximum number $m$ of increments.  In the context of a specified $m$, this paper optimizes the lengths of the $m$ possible increments to maximize throughput.

Section \ref{sec:Creating_a_bit} provides a detailed description of how we generate each bit of incremental redundancy for the NB-LDPC codes that we use.  Then, Section \ref{sec:incremental} shows that in the context of this incremental redundancy, the coding rate that first produces successful decoding is closely approximated by a normal distribution. 
Knowing a distribution that describes the coding rate of the first successful decoding facilitates optimization of the lengths of multiple-bit increments, as described in Section \ref{sec:VLFT with limited number of transmissions}.

\subsection{Creating a bit for incremental transmission}
 \label{sec:Creating_a_bit}
 
In \cite{Kasraisit2014}, Vakilinia et al. use NB-LDPC codes in a VLFT system with 1-bit increments.  
After the initial transmission, the transmitter sends one bit at a time until the decoder decodes correctly.  

Traditionally, rate-compatible codes are designed by starting
with a low-rate mother code and increasing the rate by
puncturing the code. The proposed NB-LDPC coding scheme in \cite{Kasraisit2014}
does not explicitly involve puncturing. Rather, the design starts with
a short, high-rate NB-LDPC code for which all symbols
are transmitted in the initial transmission. Each subsequent
transmission is a single bit carefully selected to help the
decoder as much as possible given its current decoding state.
The rate is gradually lowered by sending these additional bits,
each of which is a function of selected bits in the binary
representation of the non-binary symbols.

A  rate-$\frac{K}{N}$ NB-LDPC code over $GF(2^m)$ used in a binary communication link encodes an information sequence of size $Km$ bits into a sequence of size $Nm$ bits.
In order to use an NB-LDPC code with the primitive element $\alpha$ over binary-input channels, each $GF(2^m)=\{0,\alpha^0,\alpha^1,...,\alpha^{(2^m-2)}\}$ symbol is converted to $m$ bits.
 For example, consider  $GF(2^3)$ with the primitive element of $\alpha$.   Table \ref{tbl:binary_rep} shows how each element of $GF(2^3)$ can be uniquely represented in 3 bits $(g_3,g_2,g_1)$.

\begin{table}[h]
\begin{center}
  \caption{Binary representation of $GF(8)$ elements }
\begin{tabular}{ |@{{~}~}c@{~}|@{~}c@{~}|@{~}c@{~}|@{~}c@{~}|@{~}c@{~}|@{~}c@{~}|@{~}c@{~}|@{~}c@{~}|@{~}c@{~}|}
\hline $\alpha^i$&0&1&$\alpha$&$\alpha^2$&$\alpha^3$&$\alpha^4$&$\alpha^5$&$\alpha^6$\\
 \hline Poly.&0&1&$\alpha$&$\alpha^2$&$\alpha$$+$$1$&$\alpha^2$$+$$\alpha$&$\alpha^2$$+$$\alpha$$+$$1$&$\alpha^2$$+$$1$\\
 \hline $g_3$$ g_2$$ g_1$&$000$&001&010&100&011&110&111&101\\
 \hline
\end{tabular}
\label{tbl:binary_rep}
\end{center}
\end{table}

The  rate-$\frac{K}{N}$ non-binary LDPC codes proposed here initially encode a sequence of $Km$ bits ($K$ $GF(2^m)$ symbols) into a codeword of length $Nm$ bits. Through incremental redundancy, the rate is lowered from $\frac{Km}{Nm}$ to $\frac{Km}{Nm+b}$ where $b$ is number of additional incremental bits. Each additional bit is created by an XOR $(\oplus)$ combination (summation in $GF(2)$) of bits in the binary representation of one $GF(2^m)$ symbol. For each variable node, the receiver computes the reliability of each of the $2^m$$-$$1$ possible combinations of the bits in the binary representation is computed.  For example, in $GF(2^3)$ the reliabilities of the seven possible combinations $g_1,g_2,g_3,g_1 \oplus g_2,g_2 \oplus g_3,g_1 \oplus g_3, \text{and~} g_1 \oplus g_2\oplus g_3$ are computed for each variable node. Finally, the single combination bit that has the least reliability (e.g. considering all seven combinations for all variable nodes and choosing the least-reliable combination for a single variable node) is requested from the transmitter.

This is a form of active feedback in which relatively extensive feedback tells the transmitter \emph{what} to transmit in contrast to non-active feedback in which a single bit of feedback indicates {\emph{whether}} to transmit. This is a generalization of the ideas of active hypothesis testing \cite{JavidiPaper}. In \cite{Kasraisit2014} Vakilinia et al. compared the performance of a non-active feedback system and the active feedback system discussed earlier for NB-LDPC codes and showed significantly better performance with the active feedback system. The active feedback used in \cite{Kasraisit2014} tells the transmitter which bit combination to be transmitted next. This active feedback scheme does not require the receiver to transmit back the entire message, contrary to the analysis of \cite{Polyanskiy_IT_2011_NonAsym}. In the non-active feedback scheme of \cite{Kasraisit2014} the additional bits are selected at random. 

This paper considers both active and non-active feedback.  The non-active feedback in this paper corresponds to sending the XOR of all bits representing one of the variable nodes of the original rate-$k/N_0$ NB-LDPC code. This predetermined non-active feedback system performs close to the system with active feedback since the active feedback of \cite{Kasraisit2014} usually asks for the XOR of all bits for the subsequent transmissions. The figures and results in this paper  indicate whether active or non-active feedback scheme was used to generate them. 

The input frame consisting of $K$ $GF(2^m)$ information symbols is initially encoded by the rate-$\frac{K}{N}$ NB-LDPC encoder into a sequence of length $N$ $GF(2^m)$ symbols. These $GF(2^m)$ symbols are converted using their binary representations to bits. The $Nm$ bits are modulated using binary phase shift keying (BPSK) and transmitted over an AWGN channel. The additive noise is modeled as an independent, zero-mean Gaussian random sequence with variance $\sigma^2$.  
As in~\cite{Chen_Feedback_Journal_2013}, SNR is calculated as $\frac{1}{\sigma^2}$, the ratio of the transmission power to the noise variance.


\subsection{Gaussian and reciprocal-Gaussian Approximations}
 \label{sec:incremental}

\begin{figure}[t]
{\includegraphics[width=0.49\textwidth]{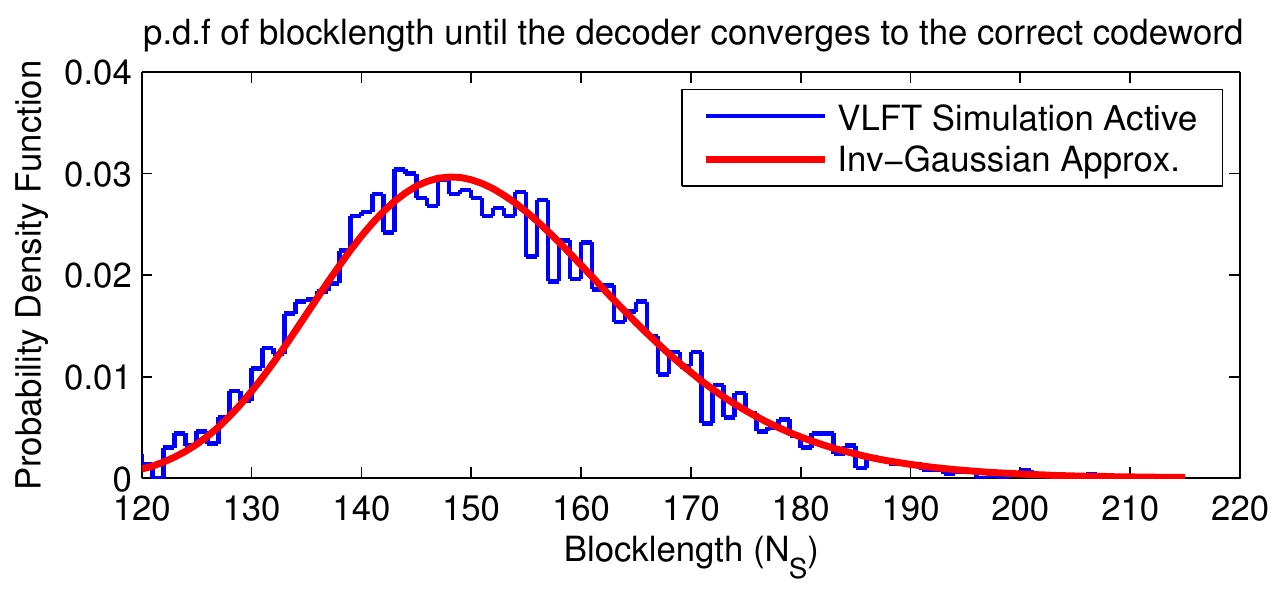}
\caption{Empirical probability mass function (p.m.f.) corresponding to the blocklength required for successful decoding for the first time in VLFT using $GF(256)$ NB-LDPC code over  SNR-2dB AWGN channel.  Also shown is the reciprocal-Gaussian approximation of \eqref{eqn:PDFNS} with $\mu_S=0.6374$ and $\sigma_S = 0.0579$. Smallest blocklength is $N_0=120$ bits with $k=96$ information bits so that the initial rate is $R_0=\frac{k}{N_0}=0.8$. }
\label{fig:NCBL}}
\vspace{0.1in}
{\includegraphics[width=0.49\textwidth]{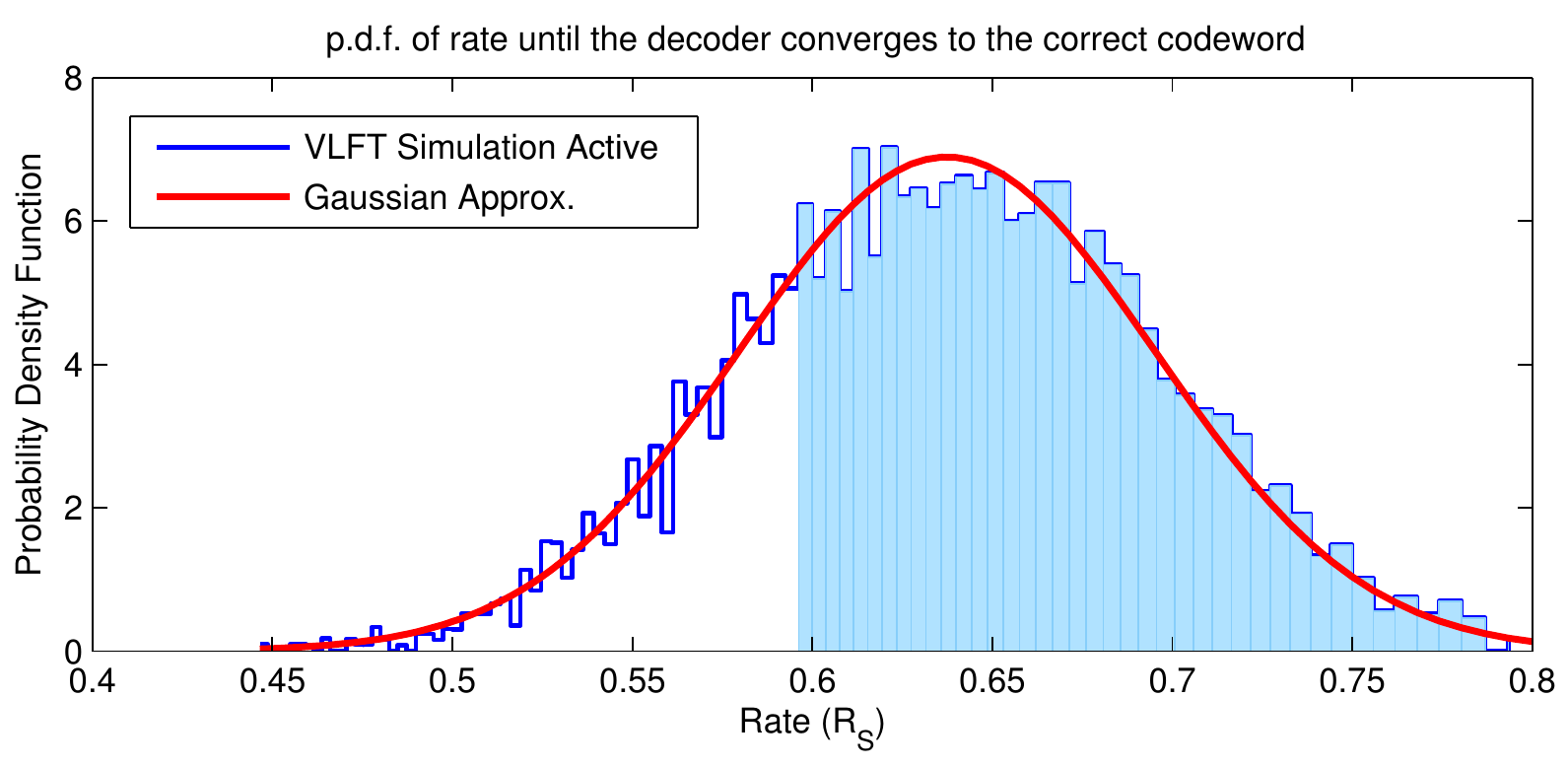}
\caption{Empirical p.m.f. corresponding to  $R_S=\frac{k}{N_S}$ computed from Fig.~\ref{fig:NCBL} and Gaussian approximation of \eqref{eqn:distribution_RS} with $\mu_S=0.6374$ and $\sigma_S = 0.0579$. }\label{fig:NCR}}
{\includegraphics[width=0.49\textwidth]{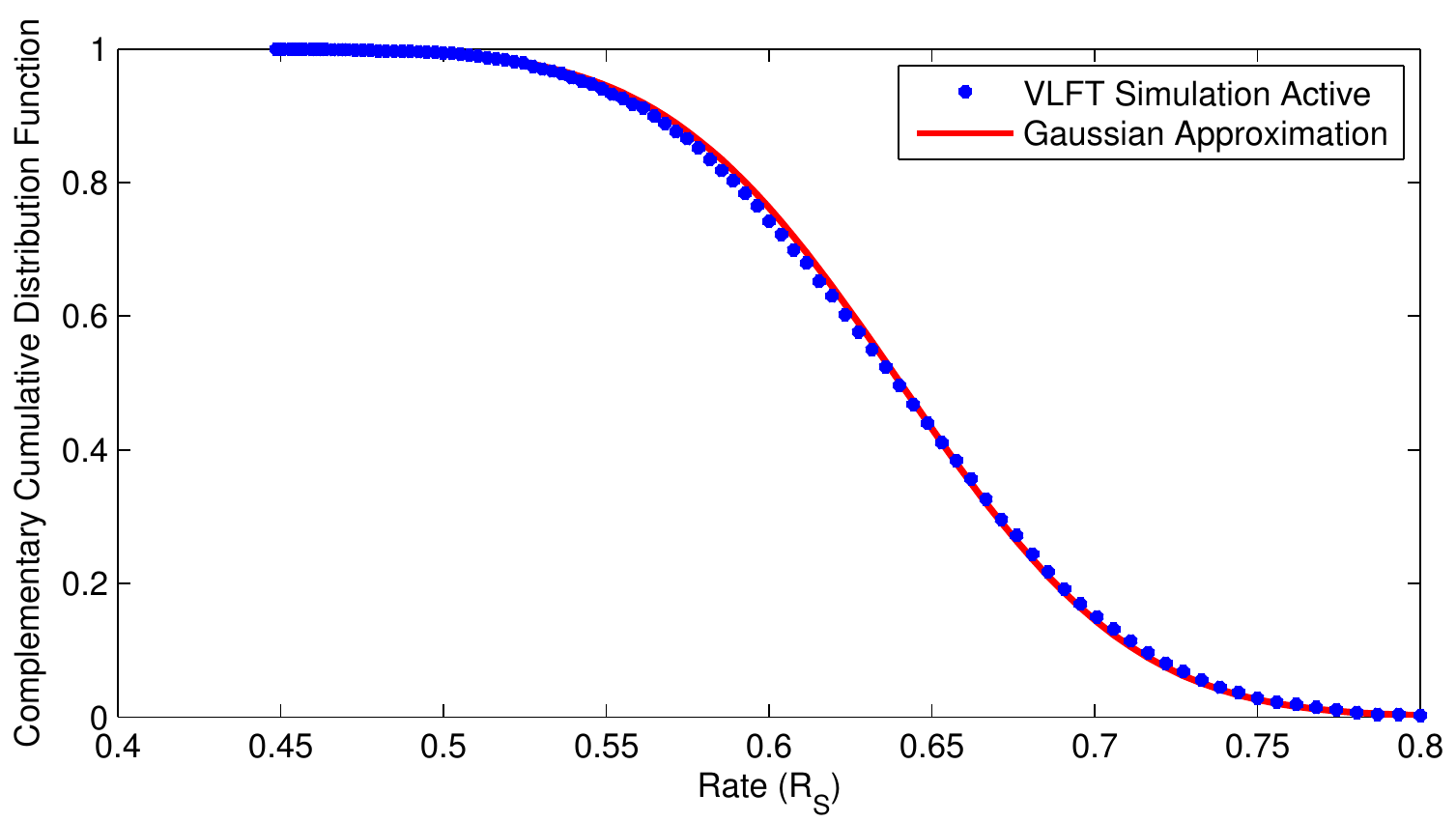}
\caption{Empirical c.c.d.f. and the approximation on the tail of a normal distribution (Q-function) corresponding to the shaded area of Fig.~\ref{fig:NCR}.}\label{fig:NCR_ccdf}}
\end{figure}

Consider a stream of incremental redundancy as described in Section \ref{sec:Creating_a_bit}  arriving one bit at a time at the receiver (after an initial transmission of a high-rate NB-LDPC code).   We are interested in the statistical behavior of the random variable describing the blocklength of the first successful decoding and the corresponding random variable describing the coding rate of that first successful decoding. 

For the system of \cite{Kasraisit2014}, the ``VLFT simulation active" plot in Fig.~\ref{fig:NCBL} shows the empirical p.m.f. of the blocklength of first successful decoding.  The total blocklength $N_S$ includes the initial block and all incremental transmissions, (with active feedback) required for receiver to decode the NB-LDPC codeword correctly for the first time. The ``VLFT simulation active" plot in Fig.~\ref{fig:NCR} shows the empirical p.m.f. of the instantaneous rate $\left(\! R_S=\frac{k}{N_S} \!\right)$ at which decoding is successful for the first time.  Fig.~\ref{fig:NCR} shows that $R_S$ is well-approximated by a normal distribution
 \begin{IEEEeqnarray}{CCC}
f_{R_S}(r)=\frac{1}{\sqrt{2\pi\sigma_S^2}}e^{-\frac{\left(\!r-\mu_S \!\right)^2}{2\sigma^2_S}}
\label{eqn:distribution_RS}
\end{IEEEeqnarray}
with mean $\mu_S=\text{E}(R_S)$ and variance $\sigma^2_S=\text{Var}(R_S)$. The intuition behind these approximations is consistent with the ``normal approximation'' of the accumulated information density due to the law of large numbers (LLN)  in \cite{Polyanskiy_CCR_2010}.

To maximize throughput, the initial code-rate of the NB-LDPC code is chosen so that almost no codeword is  successfully decoded in the initial transmission.  Thus, the empirical probability mass function (p.m.f.) of the number of additional increments required to decode correctly does not have a spike at zero.

Fig.~\ref{fig:NCR_ccdf} shows the complementary cumulative distribution function (c.c.d.f.) for the distribution of $R_S$ and the Gaussian approximation of Fig.~\ref{fig:NCR}. Fig.~\ref{fig:NCR_ccdf} confirms that the distribution of $R_S$ is well approximated by a Gaussian distribution. As discussed later, the empirical c.c.d.f is used to show that the Gaussian approximation is valid for a variety of AWGN channels including the high SNR ones using larger constellations and also for fading channels.  The ``VLFT simulation active" plot in Fig.~\ref{fig:NCR_ccdf} shows the empirical c.c.d.f. of the instantaneous rate $\left(\! R_S=\frac{k}{N_S} \!\right)$ at which decoding is successful for SNR-2dB BI-AWGN of \cite{Kasraitw2014}. This c.c.d.f. plot shows the cumulative probability that the channel supports a rate higher than the rate on the $x$ axis. This higher rate means that the decoding has been successful with a lower number of transmitted bits. The c.c.d.f. plot corresponds to the shaded area of Fig.~\ref{fig:NCR}. The ``Gaussian Approximation" plot of Fig.~\ref{fig:NCR_ccdf} corresponds to the tail probability of the standard normal distribution of Fig.~\ref{fig:NCR}.

The parameters $\mu_S$ and $\sigma^2_S$ in \eqref{eqn:distribution_RS} for a particular code need to be determined through simulation and curve fitting. Having the p.m.f. of the $N_S$, the curve fitting process involves calculating the p.m.f. and c.c.d.f. of $R_S$ and solving a linear regression problem to obtain $\mu_S$ and $\sigma_S$. 
Note that $\mu_S$ is \textit{not} the expected throughput but rather the average of the instantaneous rates supported by the channel.

The cumulative distribution function (c.d.f.) of $N_S$  is  $F_{N_S}(n)=P(N_S \leq n)$, and we have
\begin{equation}
\small
F_{N_S}(n)=P\left(\!\frac{k}{R_S}\leq n\right)\! = P\left(\!R_S \geq \frac{k}{n}\right)\! = 1-F_{R_S}(\frac{k}{n}).
\end{equation}
Taking the derivative of $F_{N_S}$ using the Gaussian approximation of $F_{R_S}$ produces the following ``reciprocal-Gaussian" approximation for p.d.f. of $N_S$:
\begin{equation}
f_{N_S}(n) = \frac{k}{n^2\sqrt{2\pi\sigma_S^2}}e^{\frac{-\left(\!\frac{k}{n}-\mu_S\right)\!^2}{2\sigma_S^2}} \,. \label{eqn:PDFNS}
\end{equation}

This p.d.f as shown in Fig. \ref{fig:NCBL} closely approximates the empirical distribution of $N_S$. For $N_1<N_2$, the probability of the decoding attempt being successful at blocklength $N_2$ but not at $N_1$ using this approximation is
\begin{IEEEeqnarray}{lCl}
\int_{N_1}^{N_2}f_{N_S}(n)dn&=&\int_{N_1}^{N_2}\frac{k}{n^2\sqrt{2\pi\sigma_S^2}}e^{\frac{-\left(\!\frac{k}{n}-\mu_S \!\right)^2}{2\sigma_S^2}}dn
\label{eqn:n1n2int}
\\
\label{eqn:CDFNS}
 &=&
Q\left(\!\frac{\frac{k}{N_2}-\mu_S}{\sigma_S}\right)\!-Q\left(\!\frac{\frac{k}{N_1}-\mu_S}{\sigma_S}\right)\! \, .
\end{IEEEeqnarray}

The increase in blocklength from  $N_1$ to $N_2$ reduces the rate from $\frac{k}{N_1}$ to $\frac{k}{N_2}$.   Note that (\ref{eqn:CDFNS}) gives the probability that the channel supports rate $\frac{k}{N_2}$ while not supporting the higher rate $\frac{k}{N_1}$. 
  The \textit{Q} functions in (\ref{eqn:CDFNS}) are due to the normally-distributed highest-rate-of-successful-decoding ($R_S$) at $\frac{k}{N_1}$ and $\frac{k}{N_2}$. 

\subsection{General Applicability of the Normal Approximation}
\label{sec:Generality}
\begin{figure}[t]
{\includegraphics[width=0.49\textwidth]{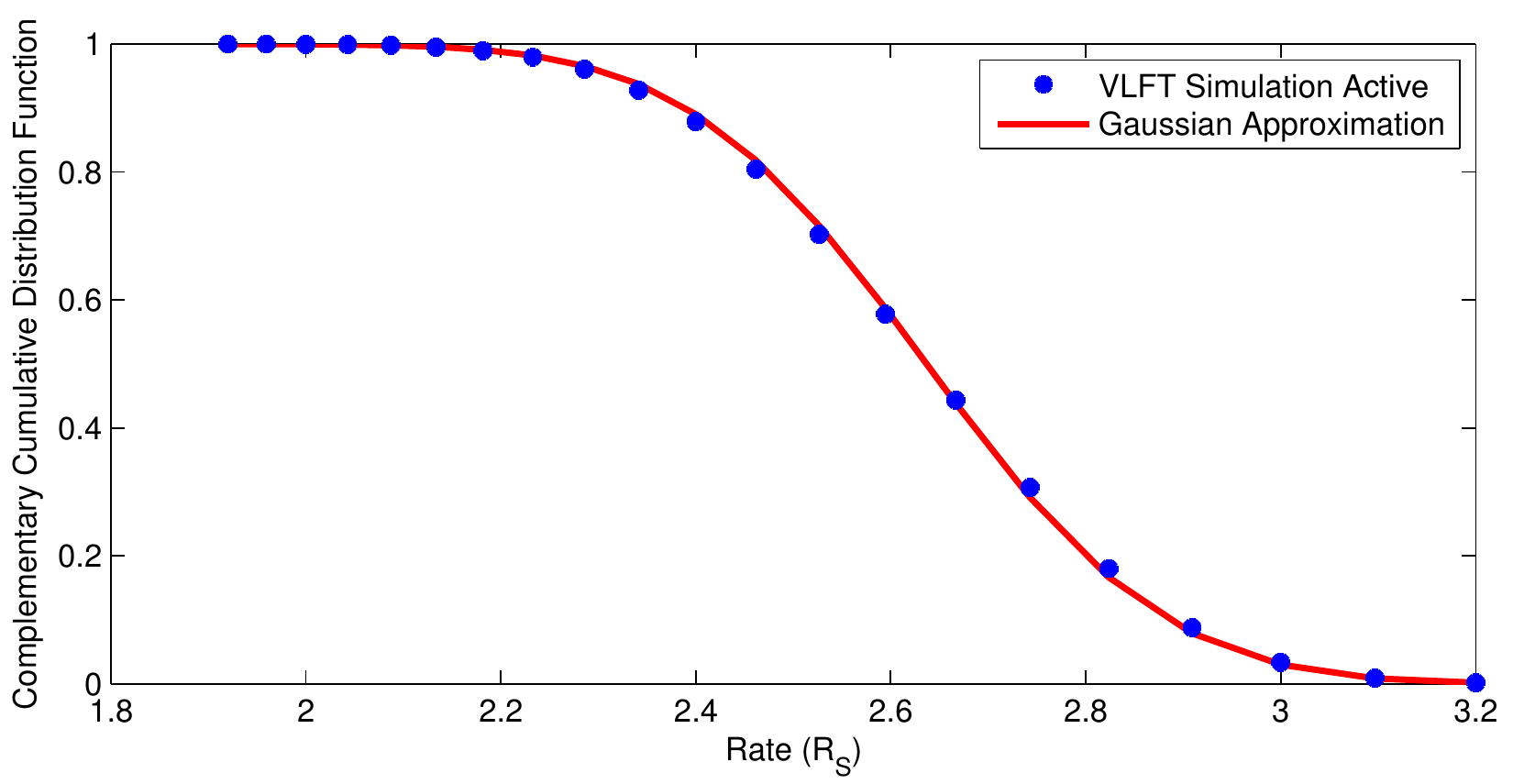}
\caption{Empirical c.c.d.f. and the approximation on the tail of a normal distribution with $\mu_S=2.63$ and $\sigma_S = 0.19$ of the instantaneous rate $\left(\! R_S=\frac{k}{N_S} \!\right)$ at which decoding is successful for SNR-8dB 16-QAM AWGN channel.}\label{fig:QAM16}}
\end{figure}

A similar Gaussian analysis is obtainable for other channels and different SNR values. Fig.~\ref{fig:QAM16} shows a similar complementary cumulative Gaussian approximation for the same $GF(256)$ NB-LDPC code of Fig.~\ref{fig:NCR} with an initial binary rate of 0.8 on SNR-8dB 16-QAM AWGN channel. Each non-binary element of the NB-LDPC code is mapped onto two 16-QAM symbols. Once again, the distribution of the instantaneous rate that the channel supports is well approximated by a normal distribution.

\begin{figure}[t]
{\includegraphics[width=0.49\textwidth]{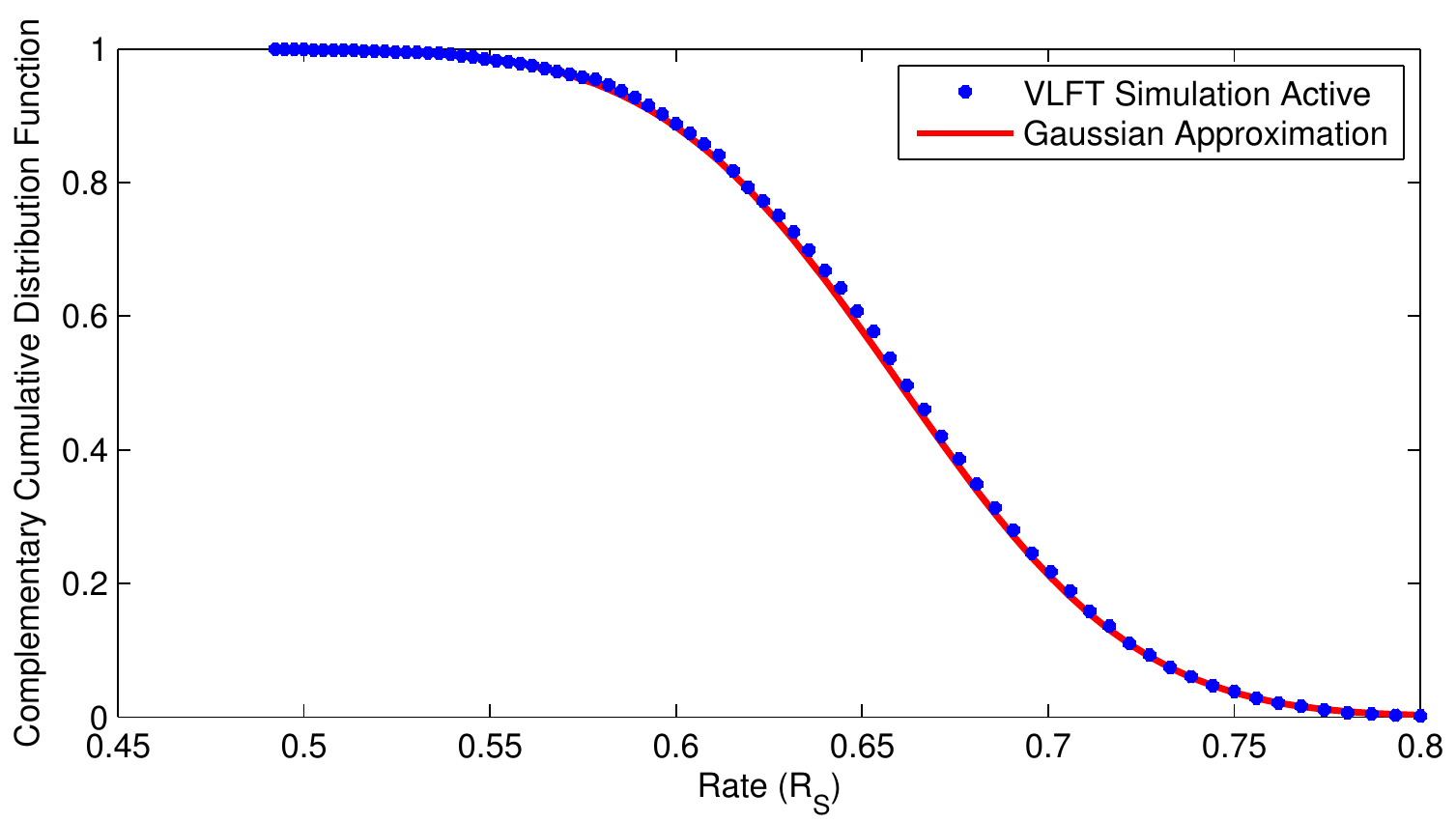}
\caption{Empirical c.c.d.f. and the approximation on the tail of a normal distribution with  $\mu_S=0.66$ and $\sigma_S = 0.05$ of the instantaneous rate $\left(\! R_S=\frac{k}{N_S} \!\right)$ at which decoding is successful for SNR-5dB AWGN fading channel.}\label{fig:WCSI}}
\end{figure}

Furthermore, Fig.~\ref{fig:WCSI} shows the complementary cumulative Gaussian approximation for the same $GF(256)$ NB-LDPC code of Fig.~\ref{fig:NCR} with an initial binary rate of 0.8 on SNR-5dB BI-AWGN fading channel with CSI knowledge at the receiver. The output of the channel, $Y=\beta X+N$  where the input $X$ is a binary phase-shift keying (BPSK) modulated signal and $N$ is the Gaussian noise with $\text{Var}(N)=\sigma^2$. The average SNR of this channel is $\frac{1}{\sigma^2}$. The coefficient $\beta$ is a Rayleigh distributed random variable satisfying $E[\beta^2]=1$. The value of $\beta$ is known at the receiver. The distribution of the instantaneous rate that the channel supports is again well approximated by a normal distribution. Since the normal distribution approximation is valid for various channels, most of the analyses in the subsequent sections of this paper are also valid for various channels with different SNR values.

To further discuss the generality of the Gaussian approximation on the rate that the channel supports in our feedback system, consider the accumulated information density $i(X,Y)$ at the receiver at the time of successful decoding. The expected value of $i(X,Y)$ is the capacity of the channel. For BI-AWGN channel, the $i(X,Y)$ is derived as follows:

{ \small \begin{IEEEeqnarray}{lll}
 i(X,Y)&=&\log_2\frac{f_{Y|X}(y|x)}{f_{Y}(y)}
\\ 
 &=&\log_2\frac{e^{-(y-x)^2/(2\sigma^2)}}{\frac{1}{2}(e^{-(y-1)^2/(2\sigma^2)}+e^{-(y+1)^2/(2\sigma^2)})}
 \\
&=&\log_2\frac{e^{-z^2/(2\sigma^2)}}{\frac{1}{2}(e^{-z^2/(2\sigma^2)}+e^{-(z+2)^2/(2\sigma^2)})}
\\
&=&1-\log_2(1+e^{-2(z+1)/\sigma^2}).
\label{eqn:VLFEN2m2}
\end{IEEEeqnarray}}

For BI-AWGN channel, $i(X,Y)$ is a function only of the noise realization $z=y-x$ for $x=\pm 1$, and hence $i(X,Y)=i(z)$. For each transmitted bit from the NB-LDPC code over the channel, there is some amount of information density accumulated. The total amount of information density accumulation ($I$) at the receiver until the receiver decodes the message correctly  

\begin{equation}
\small \label{eq:ami}
I=\sum^{N_s} _{k=1}i(z_k).
\end{equation}

The corresponding rate associated with the accumulated information density is $R_I=\frac{I}{N_s}$.  As pointed out by \cite{Polyanskiy_CCR_2010}, \eqref{eq:ami} is a sum of independent random variables for which the central limit theorem will converge quickly to  a normal distribution.  An important consideration for our approach is whether the rate at which a practical decoder succeeds also follows a normal distribution.  This hinges on the ability of a rate-compatible code family as in \cite{PBRLTCOM} to operate with a small gap from capacity over the rate range of interest. 

\begin{figure}[t]
{\includegraphics[width=0.49\textwidth]{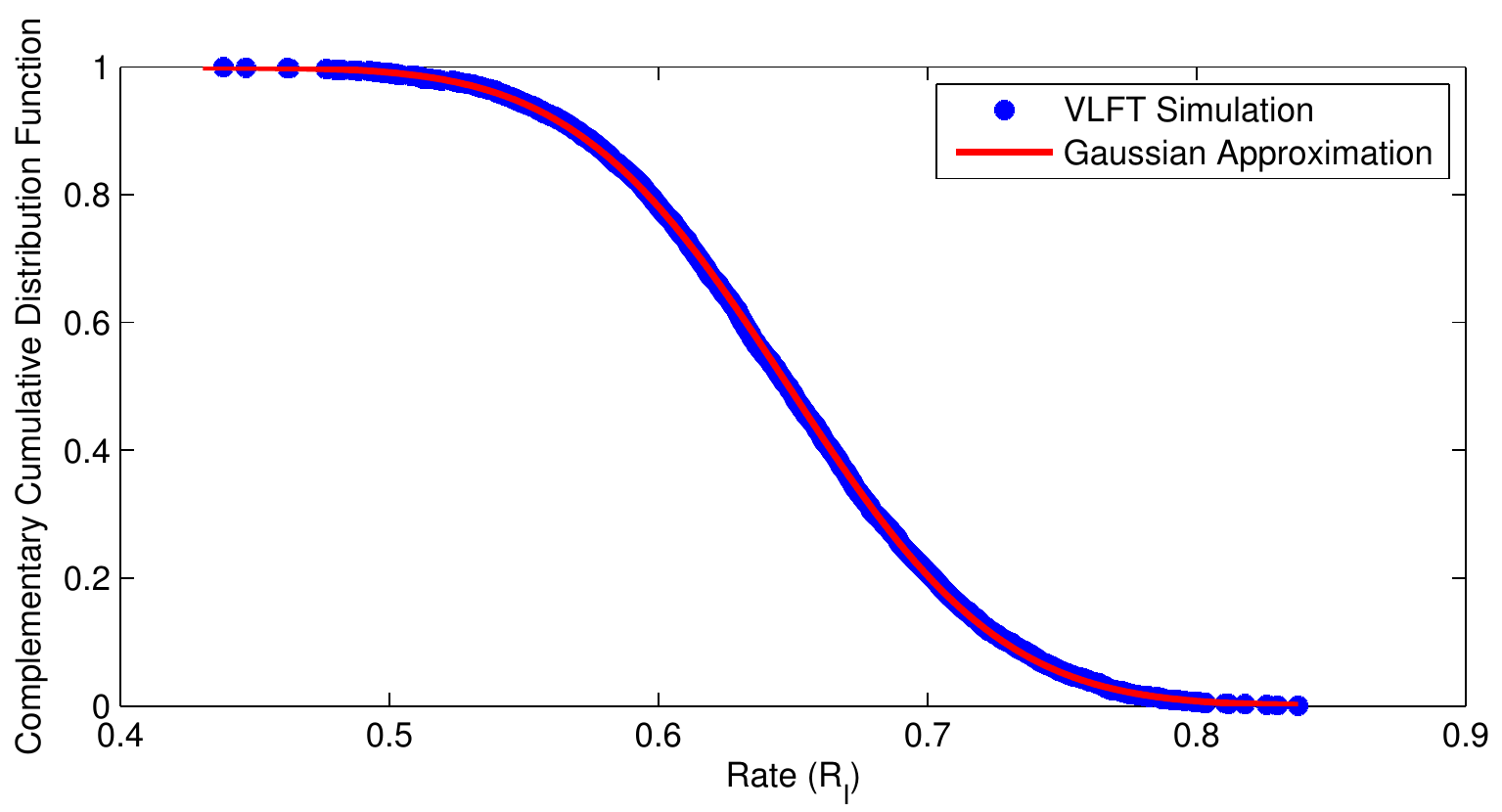}
\caption{Empirical c.c.d.f. and the approximation on the tail of a normal distribution with $\mu_S=0.64$ and $\sigma_S = 0.06$ of the average accumulated information density $\left(\! R_I=\frac{I}{N_S} \!\right)$ at which decoding is successful for SNR-2dB AWGN channel.}\label{fig:MI}}
\end{figure}

For the previously discussed SNR-2dB BI-AWGN channel,   Fig.~\ref{fig:MI} shows the c.c.d.f. of $R_I$ and the corresponding Gaussian approximation. The rate corresponding to the accumulated information density at the receiver until the decoding is successful also follows the Gaussian approximation. 

\begin{figure}[t]
{\includegraphics[width=0.49\textwidth]{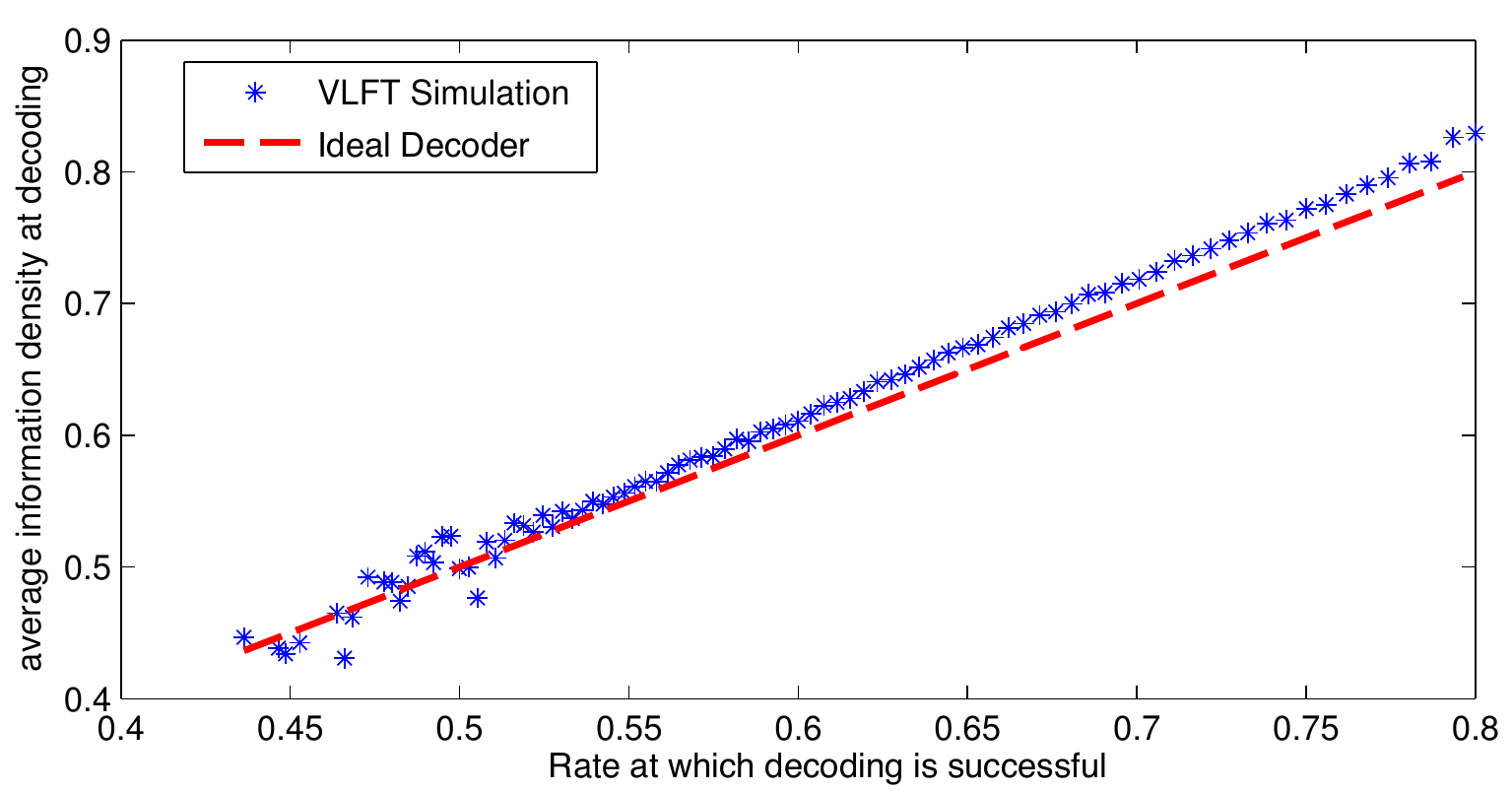}
\caption{Average amount of the accumulated information density for decoding correctly at a particular code rate for the GF(256) NB-LDPC of Fig.~\ref{fig:NCBL} code over SNR-2dB AWGN channel.}\label{fig:MI2}}
\end{figure}

Fig.~\ref{fig:MI2} shows the average accumulated information density for decoding correctly at a particular code rate for the NB-LDPC code. This figure shows on average, how much more information in number of bits the NB-LDPC code requires to decode the message correctly compared to the operating rate. The ``ideal decoder" plot in Fig.~\ref{fig:MI2} corresponds to the average accumulated information density being equal to the rate (the line of equality). 

\section{Optimizing Transmission Lengths}
\label{sec:VLFT with limited number of transmissions}


Consider the scenario in which the number of increments  (packets of incremental redundancy)  associated with a codeword that can be accumulated at the receiver is limited to $m$.  Using the p.d.f. of $N_S$ from (\ref{eqn:PDFNS}) we find the optimal blocklengths $\{N_1,N_2,\hdots,N_m\}$ to maximize the throughput.  The initial blocklength $N_1$ satisfies $N_1 \ge N_0$ where $N_0$ is the smallest possible blocklength (of the original NB-LDPC code).  Each of the additional bits beyond  $N_0$ transmitted in the first transmission is the exclusive-or of all eight bits representing one of the variable nodes of the original rate-$k/N_0$ GF(256) NB-LDPC code. 
The other transmissions use the scheme in Section~\ref{sec:Creating_a_bit} to generate the subsequent bits. 

\subsection{Throughput optimization through exhaustive search}

An accumulation cycle (AC) is a set of $m$ or fewer transmissions and decoding attempts ending when decoding is successful or when the $m^{th}$ decoding attempt fails.  If decoding is not successful after the $m^{th}$ decoding attempt, the accumulated transmissions are forgotten and the process starts over with a new transmission of the first block of $N_1$ symbols.  From a strict optimality perspective, neglecting the symbols from the previous failed AC is sub-optimal.  However, the probability of an AC failure is sufficiently small that the performance degradation is negligible.  Neglecting these symbols greatly simplifies analysis.

Define the throughput as $R_T=\frac{E[K]}{E[N]}$, where  $E[N]$ represents the expected number of channel uses in one AC and $E[K]$ is the effective number of \textit{information} bits transferred correctly over the channel in one AC. 

The expression for $E[N]$ is

{ \small \begin{IEEEeqnarray}{lCl}
E[N]&=&N_1Q\left(\!\frac{\frac{k}{N_1}-\mu_S}{\sigma_S}\!\right)
\label{eqn:EN1}
\\
&&+\sum\limits_{i=2}^m N_{i}\left[Q\left(\!\frac{\frac{k}{N_{i}}-\mu_S}{\sigma_S}\!\right)-Q\left(\!\frac{\frac{k}{N_{i-1}}-\mu_S}{\sigma_S}\!\right)\right]
\label{eqn:EN2toM}
\\
&&+N_{m}\left[1-Q\left(\!\frac{\frac{k}{N_{m}}-\mu_S}{\sigma_S}\!\right)\right].
\label{eqn:ENM}
\end{IEEEeqnarray}}

The right hand side of (\ref{eqn:EN1}) shows the contribution to expected blocklength from successful decoding on the first attempt in the AC.  $Q \left(\!\frac{\frac{k}{N_1}-\mu_S}{\sigma_S}\!\right)$ is the probability of decoding successfully with the initial block of $N_1$. Similarly, the terms in (\ref{eqn:EN2toM}) are the contributions to expected blocklength from decoding that is first successful at total blocklength $N_i$ (at the $i^{\text{th}}$ decoding attempt).  Finally, the contribution to expected blocklength from not being able to decode even at $N_m$ is $1-Q \left(\!\frac{\frac{k}{N_{m}}-\mu_S}{\sigma_S}\!\right)$ which is shown in (\ref{eqn:ENM}). Even when the decoding has not been successful at $N_m$, the channel has been used for $N_m$ channel symbols. 

The expected number of successfully transferred information bits $E[K]$ is 
\begin{equation} 
\small
E[K]=kQ \left(\! \frac{\frac{k}{N_{m}}-\mu_S}{\sigma_S} \right) \, ,
\label{eqn:EK}
\end{equation}
where $Q\left(\frac{\frac{k}{N_{m}}-\mu_S}{\sigma_S}\right)$ is the probability of successful decoding at some point in the AC.  Note that $E[K]$ depends only upon $N_m$.  In fact, for large values of $N_m$, $E[K]\approx k$ and thus not sensitive to the choice of $N_m$ 

Exhaustive search (ES) can be used to optimize  \{$N_1,N_2,\hdots,N_m$\} to maximize $R_T=\frac{E[K]}{E[N]}$. The order of complexity for ES is $O\big({N_{max}-N_0+1 \choose m}\big)$, where $N_{max}$ is the maximum allowable overall blocklength for an AC.  Since $E[K]\approx k$, maximization of $R_T$ is equivalent to minimization of $E[N]$.

\subsection{Sequential differential optimization}

Sequential differential optimization (SDO) is an extremely effective alternative to ES.  Over a range of possible $N_1$ values, SDO optimizes \{$N_2,\hdots,N_m$\} to minimize $E[N]$ for each fixed value of $N_1$ by setting derivatives to zero as follows:
\begin{equation}
\small
\left \{N_2,\hdots,N_{m} : \frac{\partial E[N]}{\partial N_i}=0, ~\forall i=1,\hdots,m\!-\!1 \right\} \, . \label{derivative0}
\end{equation}

For each $i \in \{2, \ldots, m\}$, the optimal value of $N_i$ is found by setting $\frac{\partial E[N]}{\partial N_{i-1}}=0$, yielding a sequence of relatively simple computations.  In other words, we select the $N_{i}$ that makes our previous choice of $N_{i-1}$ optimal in retrospect.  For example to find $N_2$ we compute the derivative 

\begin{equation}
\small 
\dfrac{\partial E[N]}{\partial N_1}=Q\left(\!\frac{\frac{k}{N_1}-\mu_S}{\sigma_S}\!\right)+(N_1-N_2)Q^\prime\left(\!\frac{\frac{k}{N_1}-\mu_S}{\sigma_S}\!\right)=0
\end{equation}
and solve for $N_2$ 
as 
\begin{equation}
\small
N_2=\frac{Q \left(\! \frac{\frac{k}{N_1}-\mu_S}{\sigma_S} \!\right) +N_1Q^\prime \left(\! \frac{\frac{k}{N_1}-\mu_S}{\sigma_S}\!\right)}{Q^\prime \left(\! \frac{\frac{k}{N_1}-\mu_S}{\sigma_S}\!\right)}\, ,
\label{eqn:N2D}
\end{equation}
where 
\begin{equation}
Q^\prime\left(\!\frac{\frac{k}{N_i}-\mu_S}{\sigma_S}\!\right)=\!\frac{k}{N_i^2\sigma_S}\frac{1}{\sqrt{2\pi}}e^\frac{\left(\!\frac{k}{N_i}-\mu_S\!\right)^2}{2\sigma_S^2}.
\end{equation}
%
For $i>2$,  $\frac{\partial E[N]}{\partial N_{i-1}}=0$ depends only on \{$N_{i-2},N_{i-1},N_{i}$\} as follows:
 {\small \begin{equation}
 \small 
\frac{\partial E[N]}{\partial N_{i\!-\!1}}\!=\!Q\!\left(\!\!\frac{\frac{k}{N_{i\!-\!1}}\!-\!\mu}{\sigma}\!\right)\!+\left(\!N_{i\!-\!1}\!-\!N_{i}\!\right)\!Q^\prime\!\left(\!\!\frac{\frac{k}{N_{i\!-\!1}}\!-\!\mu}{\sigma}\!\right)\!- Q\!\left(\!\!\frac{\frac{k}{N_{i\!-\!2}}\!-\!\! \mu}{\sigma}\!\right)\!.\nonumber
\end{equation}}
Thus we can solve for $N_i$ as
 \begin{equation}
 \small
 N_{i}=\frac{Q\!\left(\!\frac{\frac{k}{N_{i-1}}-\mu}{\sigma}\!\right)\!+\!N_{i-1} Q^\prime\!\left(\!\frac{\frac{k}{N_{i-1}}-\mu}{\sigma}\!\right)\!- \! Q\!\left(\!\frac{\frac{k}{N_{i-2}}-\mu}{\sigma}\!\right)}
{Q^\prime\!\left(\!\frac{\frac{k}{N_{i-1}}-\mu}{\sigma}\!\right)}\, .
\label{eqn:Ni1D}
\end{equation}


Actually, for each possible value of $N_1$, SDO can be used to produce an infinite sequence of $N_i$ values that solve  \eqref{derivative0}.  Each such sequence is an optimal sequence of increments for a given density of retransmission points on the transmission axis.  As $N_1$ increases, the density decreases.  Using SDO to compute the optimal $m$ points is equivalent to selecting the most dense SDO-optimal sequence that when truncated to $m$ points results in the highest throughput.

\subsection{Application to VLFT with $m$ transmissions}

Table~\ref{tbl:BFvsH} shows the optimized $\{N_1, N_2, \hdots, N_m\}$, resulting throughput $R_T$, and expected blocklength $\lambda=k/R_T$ for various $m$. The values obtained by SDO are very close to the values obtained by ES. 

For $m=2,5,6, \text{and } 7$, the optimized blocklengths for both approaches are the same.  For $m=3 \text{ and } 4$ the blocklengths differ only in the value of $N_m$ (shown in bold) and only by one bit.   This small difference in $N_m$ causes a negligible difference in the maximum throughput $R_T$ and minimum expected blocklength $\lambda\!=\!\frac{k}{R_T}$. Since the complexity of ES is exponential in $m$, it is infeasible to obtain a globally optimal solution for $m>7$; whereas SDO, with complexity $O(N_{max}-N_0)$, can find a solution within seconds even for large $m$. 

\begin{table}[t]
 \renewcommand*{\arraystretch}{1.2}
\begin{center}
  \caption{Optimized $\{N_1, N_2, \hdots, N_m\}$, $R_T$, and $\lambda$ from ES and SDO for $k $ = 96 bits  for VLFT on a 2 dB SNR binary-input AWGN channel using $\mu_S=0.6374$ and $\sigma_S = 0.0579$.} 
   \begin{tabular}{@{}l|c|l|l|l@{}}
Alg. & $m$ & $\{N_1, N_2, \hdots, N_m\}$ & $R_T$ & $\lambda$\\   
   \hline 
\hline 
  $\text{ES, SDO}$& 2& 158 , 188 &0.566&169.6 \\
\hline 
  $\text{ES}$& 3& 150, 167, \bf{194} &0.58638 &163.71\\
 
  $\text{SDO}$& 3& 150, 167, \bf{195} &0.58635& 163.72 \\
  \hline 
  $\text{ES}$& 4& 146, 158, 172, \bf{198}  &0.59709&160.77 \\
 
  $\text{SDO}$& 4& 146, 158, 172, \bf{197}  &0.59707& 160.78 \\
 \hline 
  $\text{ES, SDO}$& 5& 143, 153, 163, 176, 201 &0.603&159.2 \\
\hline 
$\text{ES, SDO}$& 6& 140, 149, 157, 166, 179, 204 &0.608 &157.9\\
\hline 
$\text{ES, SDO}$& 7& 139, 147, 154, 161, 170, 182, 206 & 0.611 &157.1\\
\end{tabular}
\label{tbl:BFvsH}
\end{center}
\end{table}

 \begin{figure}[t]
{\includegraphics[width=0.49\textwidth]{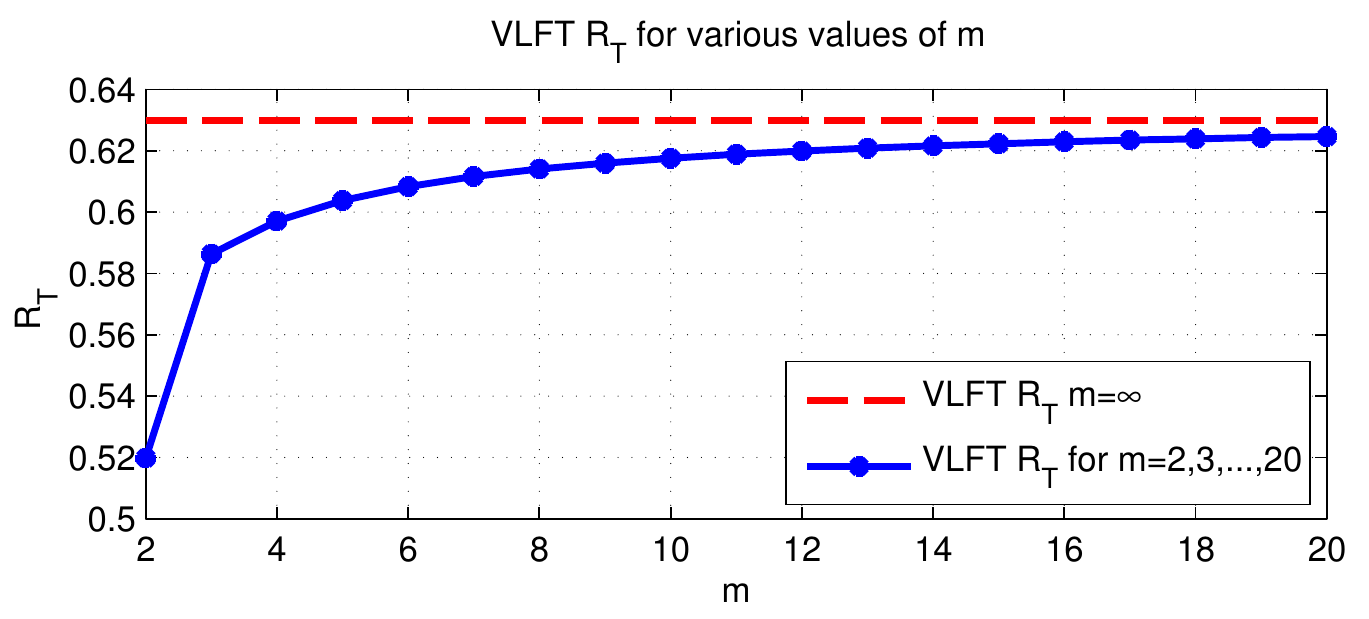}
\label{SDArt}
}
{\includegraphics[width=0.49\textwidth]{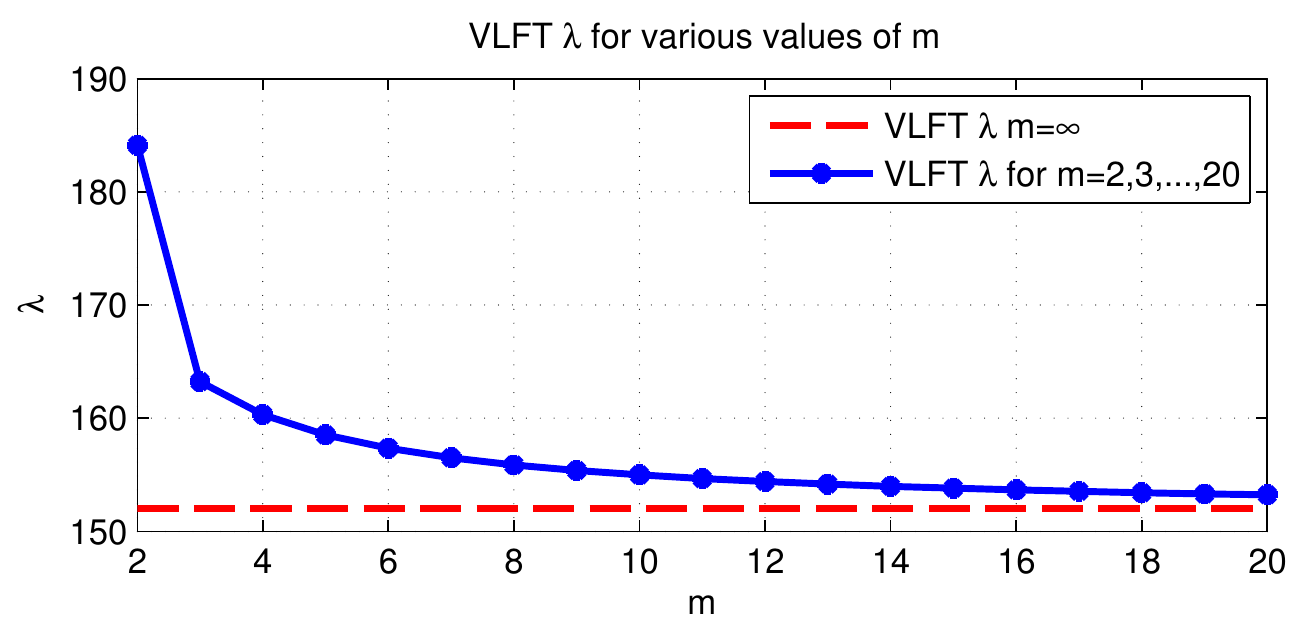}
\label{SDAlambda}}
 \caption{Throughput ($R_T$) and the expected blocklength ($\lambda$) as a function of the number of transmissions $m$ achieved by non-binary LDPC codes in the VLFT setting for $k=96$.}
 \label{fig:sdartlambda} 
\end{figure}

Fig.~\ref{fig:sdartlambda} shows the optimum $R_T$ and $\lambda$ for various $m$ using SDO. The dashed lines show the maximum achievable $R_T$ and the corresponding minimum achievable $\lambda$ with an unlimited $m$ as in \cite{Kasraisit2014}. As a function of $m$, $R_T$ quickly converges to the $m=\infty$ asymptote and even for $m \approx 10$ the throughput is close to the value achievable with an unlimited number of increments. Correspondingly, the expected latency also converges quickly and for $m \approx 10$ the expected blocklength is close to the minimum $\lambda$ achievable by unlimited transmissions of one bit at a time.

%


\section{VLF with CRC}
\label{sec:VLFCRC}
 

In this section, instead of using NTC as a genie, cyclic redundancy check (CRC) codes are used as error-detecting codes to detect whether there is an error in the decoded message. In systems incorporating CRCs, a certain number of check bits, $L_{\text{crc}}$, are computed and added to the information message of length $k_{\text{inf}}$. 

 At the receiver, the NB-LDPC decoder initially attempts to decode the received block. If decoding results in a codeword, the CRC check determines whether the check bits agree with the data by computing the checksum from the first $k_{\text{inf}}$ bits of the received sequence and comparing this checksum with the last $L_{\text{crc}}$ received bits.  In order to achieve an undetected error probability of $\epsilon$, the CRC code length $L_{\text{crc}}$ is chosen so that the overall probability of error resulting from the NB-LDPC and CRC codes combined is smaller than $\epsilon$. 

 The transmitter terminates transmission when the receiver sends feedback indicating that the decoded message passes the CRC check. If the message is correctly decoded, it passes the CRC and the transmitter moves on to the next message. If the message is decoded incorrectly and the decoded message fails to pass CRC, the transmitter sends more bits to increase reliability of the bits already transmitted. If the receiver decodes the message incorrectly and the erroneously decoded message passes the CRC check, the transmitter moves on to the next message and the packet is decoded in error. This error is undetected by the receiver.  In the case of unlimited transmissions ($m=\infty$), the transmitter transmits one bit at a time until the decoder either decodes the message correctly or until it decodes to a message that passes the CRC check. 

 \begin{figure}[t]
{\includegraphics[width=0.49\textwidth]{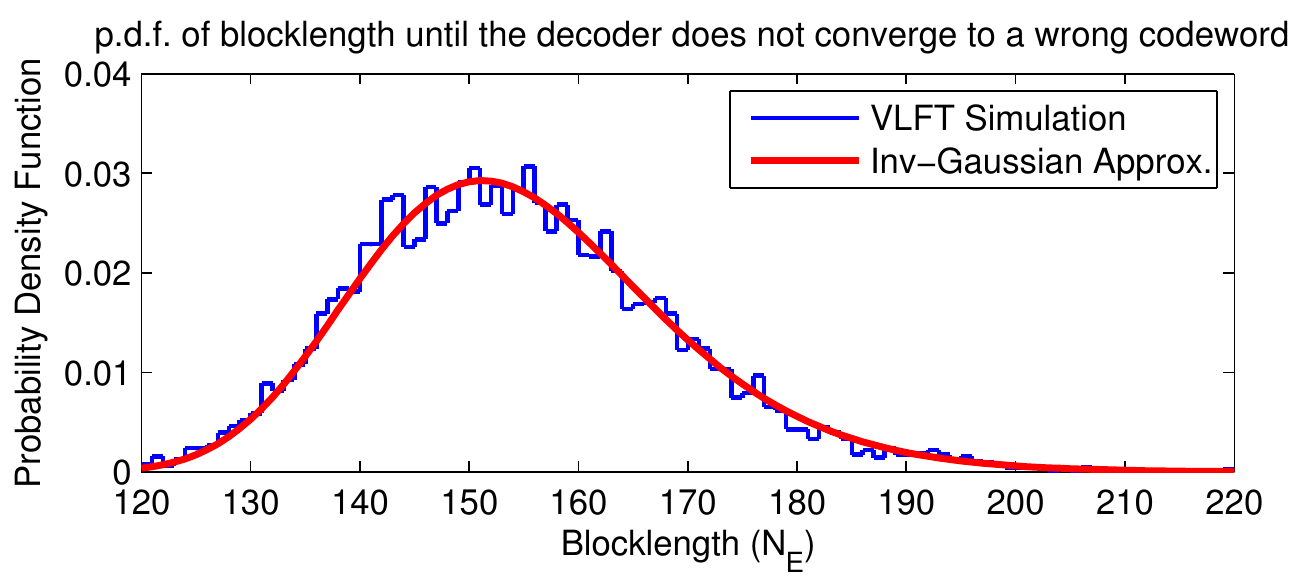}
\caption{Empirical p.m.f. and reciprocal-Gaussian fit for the shortest cumulative blocklength ($N_E$) after which decoding never again converges to an incorrect codeword. The smallest blocklength for the GF(256) NB LDPC code is $N_0=120$ bits with $k=96$ information bits. Thus, the initial rate is $R_0=\frac{k}{N_0}=0.8$. } 
\label{fig:EBL}}
\vspace{0.1in}
{\includegraphics[width=0.49\textwidth]{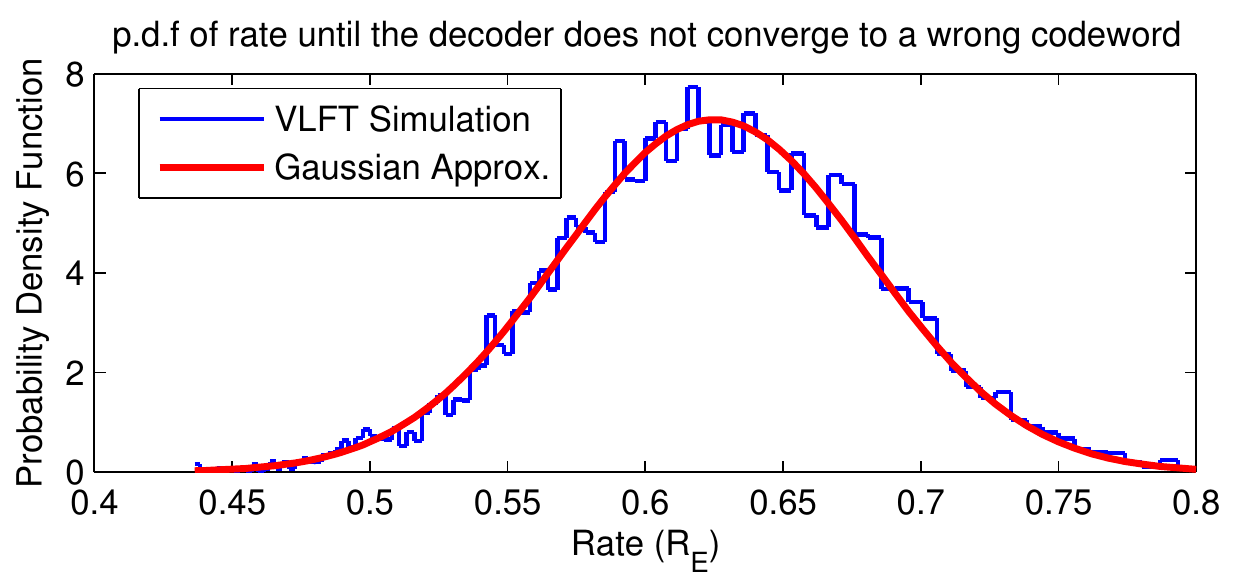}
\caption{Empirical p.m.f. and Gaussian approximation with $\mu_E=0.626$ and $\sigma_E^2=0.056 $ of $R_E$ in VLFT setting. 
}\label{fig:ER}}
\end{figure} 
 
With a limited number of transmissions, the blocklength corresponding to each transmission and the length of CRC are chosen to guarantee a probability of undetected error of at most $\epsilon$. If the message is not decoded correctly even after $m$ transmissions (and the NACKs are correctly received), the receiver deletes all received symbols and a new transmission cycle begins with the transmitter sending the original block of $N_1$ symbols.

Since the CRC as an error detection tool is used only when the decoder converges to a codeword, it is crucial to differentiate between \textit{erroneous} decoding and failure to converge to a codeword. Fig.~\ref{fig:EBL} shows the empirical p.m.f. of the required cumulative number of symbols ($N_E$) until the receiver will never again converge to an incorrect codeword.  Note that Fig.~\ref{fig:EBL} is conditioned on the decoder initially decoding to a wrong codeword at $N_0=120$.  The probability that the decoder decodes incorrectly at $N_0$ is $\gamma$. (For the experiment that produced the p.m.f. in Fig.~\ref{fig:EBL} $\gamma = 0.165$.) 

For blocklengths larger than $N_E$, the decoder either decodes correctly or fails to converge to any codeword.   This is a different condition than correct decoding, which was modeled in Figs. \ref{fig:NCBL} and \ref{fig:NCR}. Fig.~ \ref{fig:ER} shows the empirical p.m.f. of $R_E=\frac{k}{N_E}$, the instantaneous rate at which the decoder stops decoding to the wrong codeword, and the corresponding Gaussian approximation.

\begin{figure}[t]
\includegraphics[width=0.49\textwidth]{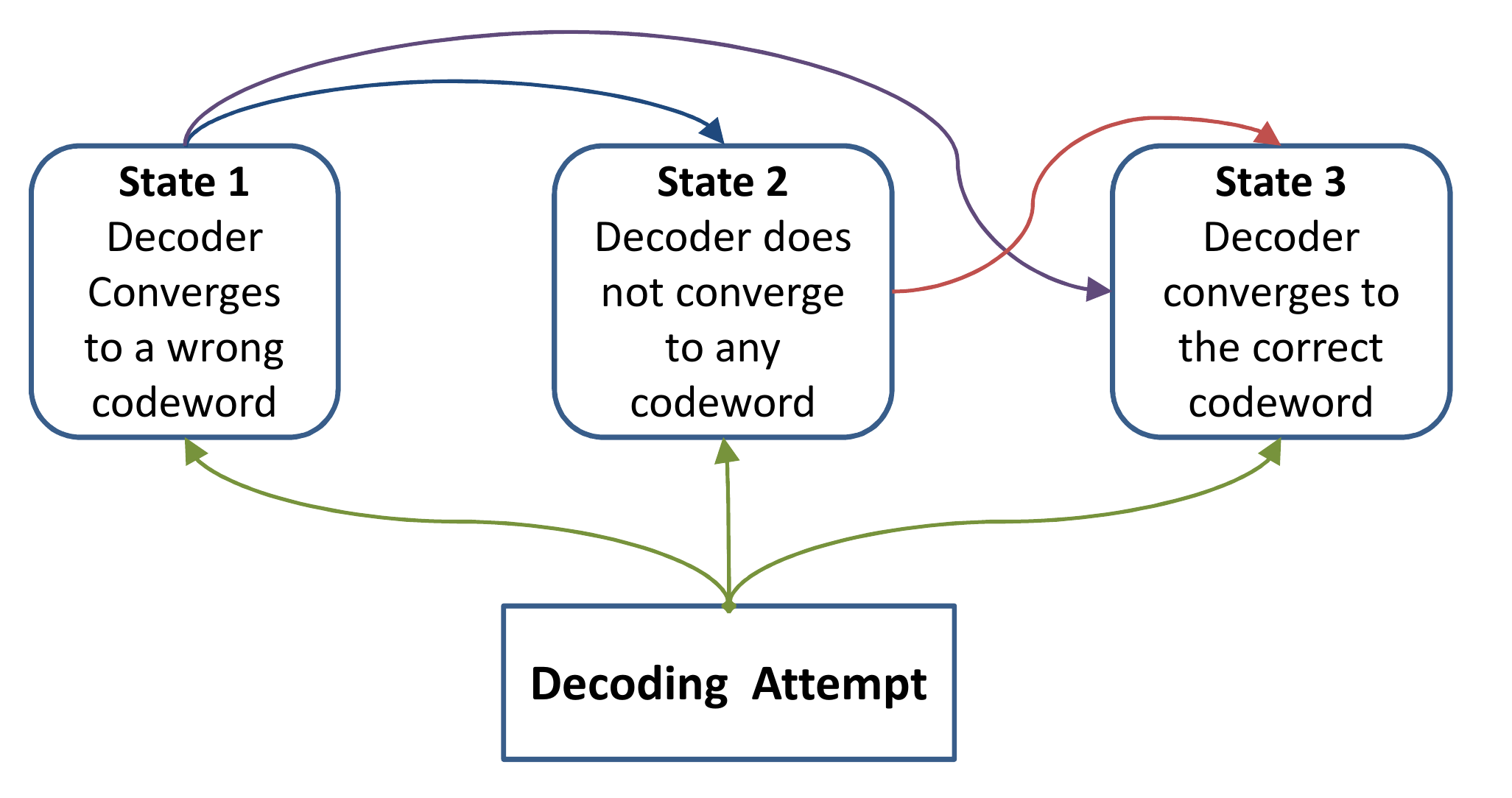}
\caption{The state diagram corresponding to LDPC coding with incremental transmissions.} 
\label{fig:statediagram}
\end{figure}

Fig.~\ref{fig:statediagram} shows the state diagram representing all the scenarios that can happen based on our simulations. 
According to our simulations, if the decoder converges to a wrong codeword, it continues to decode to the same wrong codeword even with additional incremental transmissions. The increased reliability from incremental transmissions never moves the decoder from one wrong codeword to another wrong codeword.  It only helps the decoder either to converge to the correct codeword or not to converge to any codeword at all.  Figs.~\ref{fig:NCBL}, \ref{fig:NCR} correspond to the blocklength and rate of entry to state 3.  Figs.~\ref{fig:EBL}, \ref{fig:ER} correspond to the blocklength and rate of leaving state 1.

In this section, similar to the case of $m=\infty$ VLFT, the transmitter sends one bit of incremental redundancy at a time until the decoder converges to the correct codeword or converges to an incorrect codeword that passes the CRC check. We require an undetected error probability of smaller than $\epsilon$. If the transmission starts with a blocklength of length $N_0$, the total probability of error is $\gamma \times 2^{-L_{\text{crc}}}$, where $2^{-L_{\text{crc}}}$ is approximately the probability of error that the CRC checks for a wrong codeword. 
This paper uses standard CRC codes. However, for the best error detection, the CRC codes can be designed specifically for a particular code as shown in \cite{Chung-Yu_tcom}.  

For the error probability constraint of $\epsilon$, we choose the length of the CRC code so that $\gamma \times 2^{-L_{\text{crc}}} < \epsilon$. For example, if $\epsilon$ is set to be $10^{-3}$ and $\gamma = 0.165$, the length of the CRC code $L_{\text{crc}}=8$ is required to guarantee the overall probability of error, $\gamma \times 2^{-L_{\text{crc}}} = 6.25 \times 10^{-4} < \epsilon = 10^{-3}$.

As will be illustrated in the results section (Section  \ref{sec:results}), the throughput of this scheme can be well predicted by the results obtained from VLFT with unlimited transmissions (Section  \ref{sec:VLFT with limited number of transmissions}) modified by a factor of $\frac{k-L_{\text{crc}}}{k}$ that captures the back-off in rate due to the CRC overhead.  For example, in our previous analysis from Table \ref{tbl:BFvsH} for $m=\infty$, the rate is 0.632 while with a CRC of length 8, for $k_{\text{inf}}=96-8 = 88$ the rate is predicted to be $\frac{96-8}{96} \times 0.632 = 0.579$. As the simulation results of Section~\ref{sec:results} show, the actual achieved rate is $0.575$ with an undetected error probability of $8.04 \times 10^{-4}$. We will discuss these results in more detail in Section \ref{sec:results}.

\section{VLF with CRC and Limited Transmissions}
\label{sec:VLFCRC_limited}
 
In VLF with a limited number of transmissions, the length of each incremental transmission should be selected to maximize
$R_T=\frac{E[K| L_{\text{crc}}]}{E[N]}$, where $E[N]$ is given by \eqref{eqn:EN1} and $E[K|L_{\text{crc}}]$ is the effective number of transmitted information bits, computed as
 \begin{equation} 
\small
E[K|L_{\text{crc}}]=(K-L_{\text{crc}})\left[Q \left(\! \frac{\frac{K}{N_{m}}-\mu_S}{\sigma_S} \right)-P_{N_1} 2^{-L_{\text{crc}}}~ \right],
\label{eqn:EK}
\end{equation} 
under the constraint that the probability of undetected error $P_{N_1} ~ 2^{-L_{\text{crc}}}<\epsilon$. $P_{N_1}$ is the probability of converging to an incorrect codeword at blocklength $N_1$.

An approximation technique similar to the one used in optimizing the length of each incremental redundancy block in VLFT is used here: $ \left[Q \left(\! \frac{\frac{k}{N_{m}}-\mu_S}{\sigma_S} \right)-P_{N_1} ~ 2^{-L_{\text{crc}}}~ \right] \approx 1$. The optimization problem of maximizing $R_T=\frac{E[K| L_{\text{crc}}]}{E[N]}$ reduces to minimizing $E[N]$ for each $L_{\text{crc}}$. The SDO technique used in Section \ref{sec:VLFT with limited number of transmissions} can be used here under the additional constraint that $P_{N_1} ~ 2^{-L_{\text{crc}}}<\epsilon$. 

\begin{table}[t]
 \renewcommand*{\arraystretch}{1.13}
\begin{center}
  \caption{Optimized $\{N_1,\hdots,N_m\}$ for $m$$=$$5$ in VLF-with-CRC using SDO for different values of $L_{\text{crc}}$.  The exact same values were obtained by ES.}
   \begin{tabular}{c|c|c|c|c}
$L_{\text{crc}}$ & $\{N_1, N_2, \hdots, N_5\}$ & $\lambda$& $R_T$& $\epsilon$ \\
\hline 
1 & 193, 198, 205, 216, 241 & 193.27 & 0.49 & $8.95 \times 10^{-4}$\\
2 & 187, 192, 199, 210, 235 & 187.38 & 0.50 & $9.02  \times 10^{-4} $
 \\
3 & 180, 185, 192, 203, 228 & 180.67 & 0.51 & 9.82 $ \times 10^{-4} $\\
4 & 174, 180, 187, 198, 222 & 175.14 & 0.52 & $9.14  \times 10^{-4}$ \\
5 & 166, 172, 180, 192, 216 & 168.48 & 0.54 & $9.62  \times 10^{-4}$ \\
6 & 157, 164, 172, 184, 209 & 162.68 & 0.55 & $9.58  \times 10^{-4}$ \\
7 & 143, 153, 163, 176, 201 & 159.14 & 0.56 & $9.44  \times 10^{-4}$ \\
8 & 143, 153, 163, 176, 201 & 159.07 & 0.55 &$ 4.72  \times 10^{-4}$ \\
9 & 143, 153, 163, 176, 201 & 159.04 & 0.54 &$ 2.36  \times 10^{-4}$ \\
10 & 143, 153, 163, 176, 201 & 159.02 & 0.54 & $1.18  \times 10^{-4}$
\end{tabular}
\label{tbl:VLFCRCN1N5}
\end{center}
\vspace{-0.1in}
\end{table}

For each $L_{\text{crc}}$, the optimized $\{N_1,\hdots,N_m\}$ values for this case are identical for SDO and ES and the values are given in Table \ref{tbl:VLFCRCN1N5}.  For small values of $L_{\text{crc}}$ we need to use a large value of $N_1$ to make sure $P_{N_1} ~ 2^{-L_{\text{crc}}}<\epsilon$. As a larger value of $L_{\text{crc}}$ is selected, $N_1$ and consequently $\{N_2,\hdots,N_5\}$ decrease while the error probability constraint is still satisfied. For $L_{\text{crc}}=7$ the set of  $\{N_1,\hdots,N_5\}=\{143, 153, 163, 176, 201\}$ minimizes the expected latency $\lambda$ and maximizes $R_T$. For larger values of $L_{\text{crc}}>7$, the set of optimum blocklengths does not change and only the overall probability of error decreases as the CRC length is increased.

The optimal set of blocklengths for $L_{\text{crc}} \ge 7$ and $m=5$ is the same as the set for VLFT and $m=5$ from Table \ref{tbl:BFvsH}. The intuition for this is that once $L_{\text{crc}}$ is large enough that decoding decisions are extremely reliable, the optimal blocklengths for VLF-with-CRC should match those of VLFT.  Because the blocklengths are identical, the throughput $R_T$ for $m=5$ with $L_{\text{crc}}=7$ can be computed by reducing the $R_T$ in Table  \ref{tbl:BFvsH} to account for the overhead of the CRC.  The reduction from the $m=5$ VLFT rate $R_T =0.603$ is $\frac{96-7}{96}$ where $\frac{96-7}{96} \times 0.603 = 0.559$ which corresponds to the $R_T$ from Table~\ref{tbl:VLFCRCN1N5} for $L_{\text{crc}}=7$.  
While both SDO and ES give the same values for different $L_{\text{crc}}$ values, the order of complexity for SDO is $O(L_{\text{crc}}(N_{max}-N_0))$ while with ES algorithm the complexity has the much larger order of $O\big(L_{\text{crc}}{N_{max}-N_0 \choose m}\big)$. As the simulation results of Section~\ref{sec:results} show, the actual achieved rate is $0.541$ with an undetected error probability of $5.75 \times 10^{-4}$. 


For the simulations in Section~\ref{sec:results} the CRC code used for $L_{\text{crc}}=7$ has a polynomial representation of 0x09 ($x^7  + x^3 +1$). This CRC code has been used by Telecommunication Standardization Sector of the International Telecommunications (CCITT) which sets international communications standards. The CRC code used for $L_{\text{crc}}=8$ has a polynomial representation of 0x07 ($x^8  + x^2  + x + 1$) and is used in MultiMedia Cards (MMC) and Secure Digital (SD) cards.

\section{Two-phase VLF}
\label{sec:VLFoptimization}
Now we consider the two-phase VLF model in which the transmitter (source) uses the primary communication channel to confirm whether the receiver (destination) has decoded to the correct codeword. As in \cite{firing_genie}, the two-phase incremental redundancy scheme has a \textit{communication} phase followed by a \textit{confirmation} phase.  


\begin{figure}[t]
\includegraphics[width=0.49\textwidth]{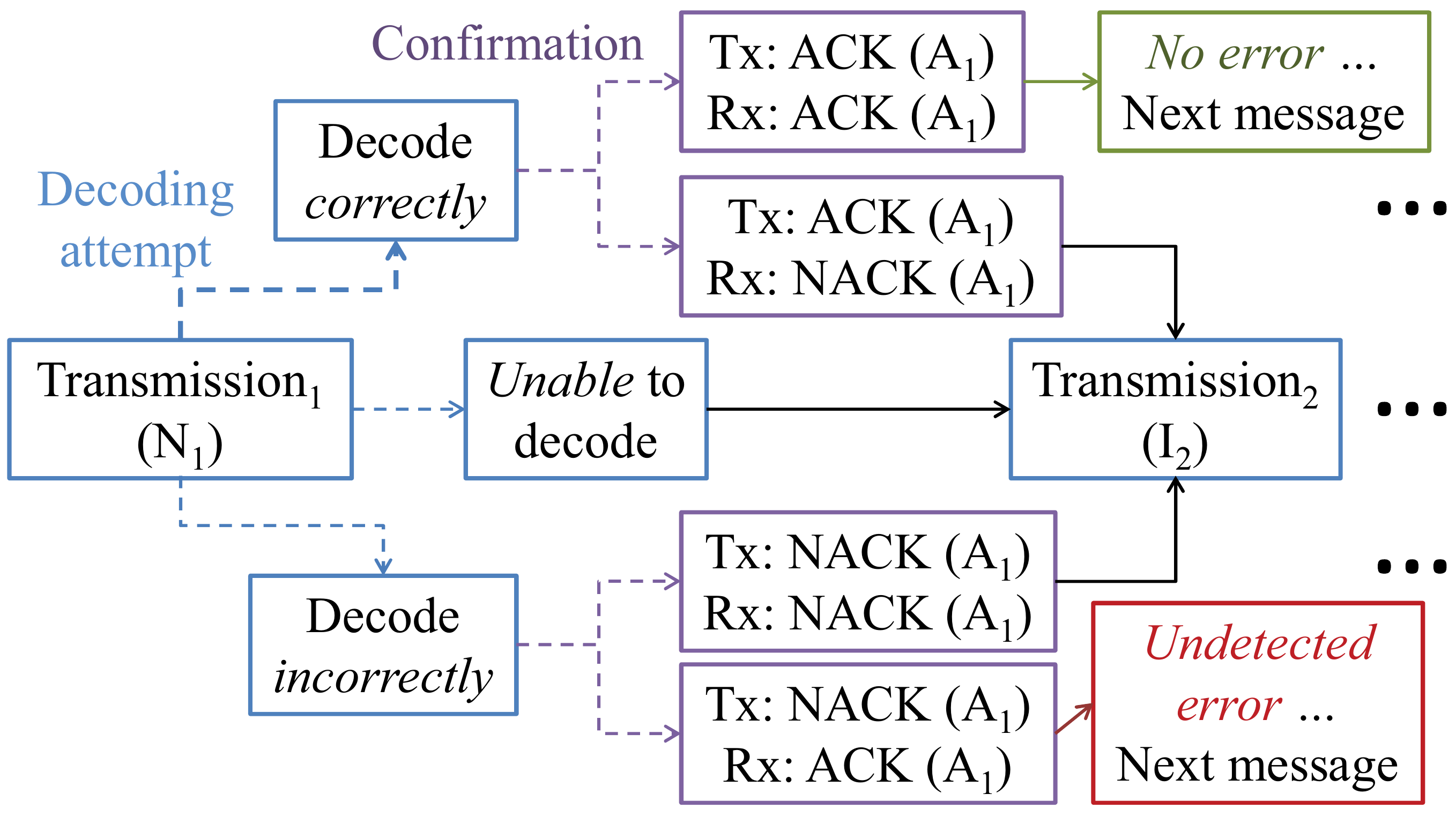}
\caption{Two-phase VLF block diagram and the forward transmission stages in two-phase VLF systems.} 
\label{fig:tree}
\end{figure}

Fig.~\ref{fig:tree} shows a block diagram for the two-phase communication scheme. Starting at the left, a message block of size $N_1$ is transmitted (communication phase).
If the destination decodes correctly, the source sends a coded forward ``ACK" on the same forward noisy channel to confirm the successful decoding (confirmation phase). If the destination decodes incorrectly, the source sends a coded forward NACK. The ACKs and NACKs are repetition codes of length $A_1$ symbols and are transmitted over the same forward noisy channel from the transmitter (source) to the receiver (destination).
If the decoder does not converge to any codeword with $N_1$ symbols, the transmitter skips the unnecessary confirmation phase and immediately transmits the second increment of $N_2-N_1$ bits.

In the two-phase VLF setting, we use the probability distributions of $N_S$, $R_S$, $N_E$ and $R_E$ from Figs.~\ref{fig:NCBL}, \ref{fig:NCR}, \ref{fig:EBL}, and \ref{fig:ER}. The optimization problem is to maximize $R_T=\frac{E[K]}{E[N]}$ where 

\begin{equation}
 \small{
 E[K]=k\left(\sum\limits_{i=1}^{m} P^{SS}_i\right)\approx k\left(\!Q\left(\!\frac{\frac{k}{N_m}\!-\!\mu_S}{\sigma_S}\!\right)-\!\sum\limits_{i=1}^{m-1} P^{EE}_i\!\right)\, ,
\label{eqn:VLFEK}}
 \end{equation}
with $P_i^{EE}$ representing the probability the receiver decodes both the message and the NACK erroneously and $P_i^{SS}$ is the probability the receiver decodes both message and ACK successfully.   Note that \eqref{eqn:VLFEK} assumes (consistent with our simulation results) that once the decoder is in state 3 of Fig. \ref{fig:statediagram}, it does not return to state 1 even if a forward ACK is incorrectly received as a forward NACK.  In any case, as in Section \ref{sec:VLFTNBLDPC} we assume  $E[K]\approx k$.  

The expected number of symbols transmitted in an AC is
 \begin{align}{\small
 E[N] =& \sum\limits_{i=1}^m(N_i\!+\!A_i)\left[P^{SS}_i\!+\!P^{EE}_i\right]\!+\!A_i\left[P^{SE}_i\!+\!P^{ES}_i\right] \label{eqn:VLFEN2m21}\\
&\!{+N_m \left(\! 1-\left(\sum\limits_{i=1}^{m} P^{SS}_i+\sum\limits_{i=1}^{m} P^{EE}_i\right)\right)}  \label{eqn:VLFENm} \, ,}\end{align}
where $P_i^{SE}$  is the probability of decoding the message successfully but decoding the ACK as a NACK. Conversely, $P_i^{ES}$ is the probability of decoding the message erroneously but decoding the NACK successfully. The term multiplying $N_m$ in (\ref{eqn:VLFENm}) is the probability that an AC ends without satisfying either of the stopping conditions. (\ref{eqn:VLFENm}) is also approximated to $N_m \left(\! 1- Q\left(\!\frac{\frac{k}{N_m}\!-\!\mu_S}{\sigma_S}\!\right)+ P^{SE}_m\!\right)$. The probabilities $P_i^{SS}$, $P_i^{EE}$, $P_i^{SE}$, and $P_i^{ES}$ are computed as follows:

 { \small \begin{IEEEeqnarray}{lll}
P^{SS}_i&=&\!\left[Q\!\left(\!\! \frac{\frac{k}{N_i}\!-\!\mu_S}{\sigma_S}\!\right)\!-\!Q\!\left(\!\frac{\frac{k}{N_{i\!-\!1}}\!-\!\mu_S\!}{\sigma_S}\!\right)\!\right]
\left[1\!-\!Q\!\left(\!\frac{\sqrt{A_i}}{\sigma_c}\!\right)\!\right]
\label{eqn:VLFEN2m1}
\\
P^{EE}_i&=&\,\left[\gamma\left(\! 1-Q\! \left(\! \frac{\frac{k}{N_i}\!-\!\mu_E}{\sigma_E}\!\right)\!\right)\right]\left[Q\!\left(\!\frac{\sqrt{A_i}}{\sigma_c}\!\right)\right] \label{eq:PEE}
\\
P^{SE}_i&=&\!\left[Q\left(\!\frac{\frac{k}{N_i}\!-\!\mu_S}{\sigma_S}\!\!\right)\!-\!Q\left(\!\frac{\frac{k}{N_{i-1}}\!-\!\mu_S}{\sigma_S}\!\right)\right]
\left[Q\left(\!\frac{\sqrt{A_i}}{\sigma_c}\!\right)\right]
\\
P^{ES}&=&\left[\gamma\left(\!1-Q\left(\!\frac{\frac{k}{N_i}\!-\!\mu_E}{\sigma_E}\!\right)\!\right)\right]\left[\left(\!1\!-\!Q\!\left(\!\frac{\sqrt{A_i}}{\sigma_c}\!\right)\right)\right] \, .
\label{eqn:VLFEN2m2}
\end{IEEEeqnarray}}

In \eqref{eqn:VLFEN2m1} the probability of decoding correctly at $N_i$ and not at blocklengths smaller than or equal to  $N_{i-1}$ is $Q\left(\!\frac{\frac{k}{N_i}-\mu_S}{\sigma_S}\!\right)-Q\left(\!\frac{\frac{k}{N_{i-1}}-\mu_S}{\sigma_S}\!\right)$ and $Q\left(\!\frac{\sqrt{A_1}}{\sigma_c}\!\right)$ is the probability that the  ACK is decoded as a NACK, where $\sigma_c$ is the standard deviation of the channel noise.  In \eqref{eq:PEE}, $\gamma\left[1-Q\left(\!\frac{\frac{k}{N_i}\!-\!\mu_E}{\sigma_E}\!\right)\right]$ is the probability of decoding erroneously at $N_i$.     

We optimize the blocklengths for two-phase VLF to  maximize $R_T$ under the constraint that
$
\sum\limits_{i=1}^mP^{EE}_i<\epsilon \, ,
$
using both ES and SDO approaches from Section \ref{sec:VLFT with limited number of transmissions} for fixed values of $\{A_1,\hdots,A_m\}$. For ES we considered values of $N_1\leq N_2\leq \dots \leq N_m$ and constrained $N_m$ to be no larger than the blocklength corresponding to a rate-0.1 code ($N_m \leq 10k$).  For SDO  we considered $N_1$ values ranging from the initial coding length $N_0$ to $3k$, which was the range that gave useful values of $\epsilon$.  

Table \ref{tbl:SDAESVLF} shows two sets of $\{N_1,\hdots,N_m\}$ with $m=5$ obtained for different $N_1$ in SDO with $\epsilon \!\approx\! 10^{-3}$. The optimized $\{N_1,\hdots,N_m\}$ with $\epsilon\!\leq\!10^{-3}$ from ES is close to the SDO optimized blocklengths. The optimized blocklengths from SDO can also be used as optimization limits for ES algorithm and significantly reduce the ES optimization space.   

\begin{table}[b]
 \renewcommand*{\arraystretch}{1.13}
\begin{center}
  \caption{Optimized $\{N_1,\hdots,N_m\}$ for $m$$=$$5$ two-phase VLF using SDO and ES with $\{A_1,\hdots,A_5\}=\{5,4,3,3,3\}$.}
   \begin{tabular}{c|c|c|c|c|c}
Alg. & $k$ & $\{N_1, N_2, \hdots, N_5\}$ & $\lambda$ & $R_T$ & $\epsilon$ \\
\hline 
SDO & 96& 145, 156, 167, 180, 202 & 166.1 & 0.5779 & 1.2E-3\\
SDO & 96& 146, 158, 171, 188, 230 & 166.6 & 0.5762 & 9.4E-4\\
\hline 
ES & 96 & 146, 158, 170, 184, 211 & 166.4 & 0.5771 & 9.9E-4 \\
\end{tabular}
\label{tbl:SDAESVLF}
\end{center}
\vspace{-0.1in}
\end{table}


\section{Results}
\label{sec:results}

\begin{figure}[t]
{\includegraphics[width=0.49\textwidth]{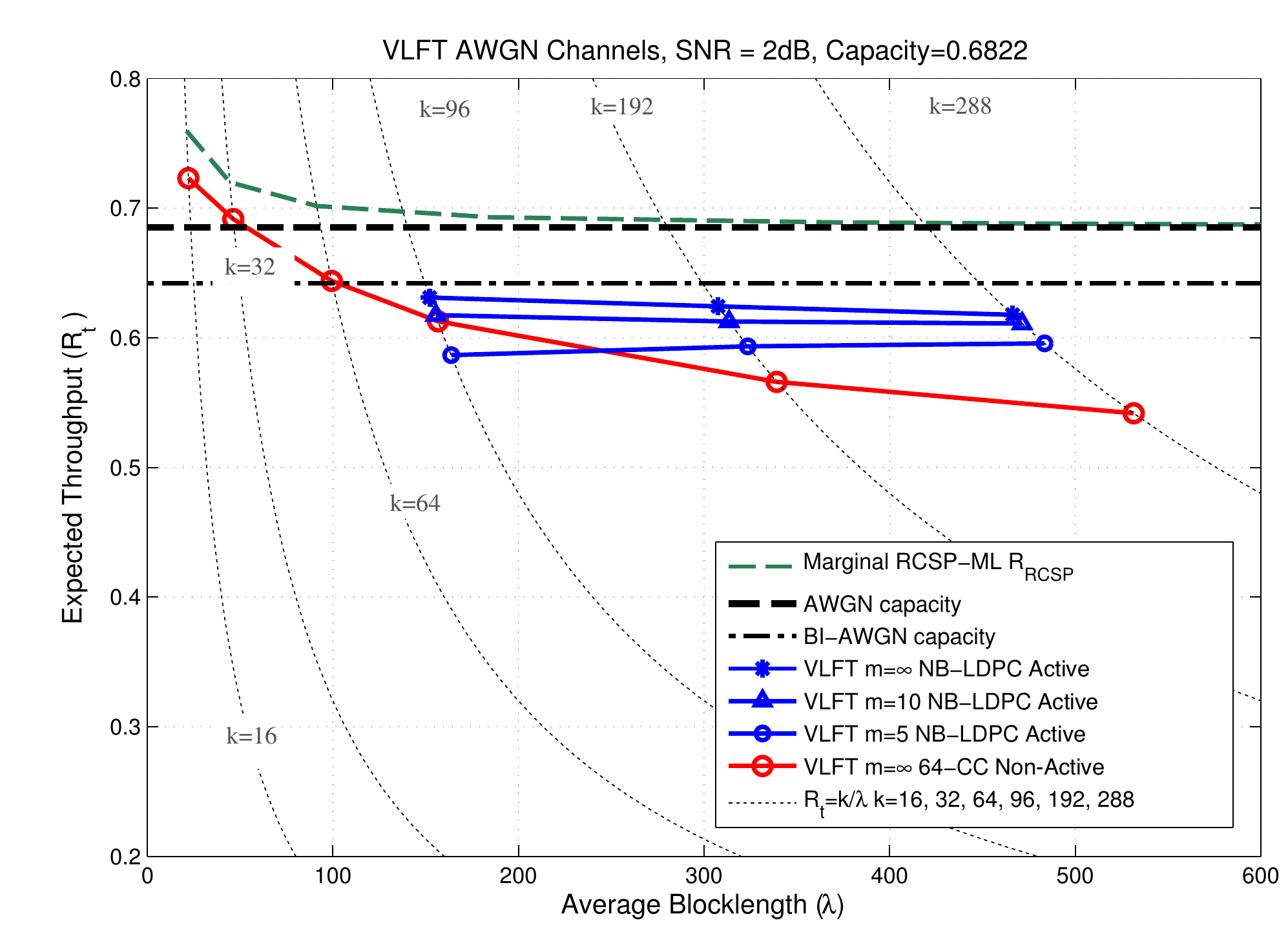}
\caption{$R_T$ vs. $\lambda$ for NB-LDPC and 1024-state convolutional codes for VLFT with $m=\infty$, $m=10$, and $m=5$.
}\label{fig:VLFT_rtlambda}}
\end{figure}

Fig.~\ref{fig:VLFT_rtlambda} shows $R_T$ versus $\lambda$ for NB-LDPC and convolutional codes using VLFT. In VLFT with an unlimited number of transmissions (1-bit increments), convolutional codes with ML decoders perform very well at short average blocklengths of up to 100 bits.  VLFT schemes have throughputs greater than capacity at short blocklengths because of the NTC. Convolutional codes follow the marginal RCSP-ML (with unconstrained input) plot closely at short-blocklength with a small gap that is due to the binary input for convolutional codes. At longer blocklengths of about 200 bits, marginal RCSP-ML rate approaches the capacity. NB-LDPC codes outperform convolutional codes at longer blocklengths because the codeword error rate of convolutional codes increases once the blocklength exceeds twice the traceback depth \cite{Anderson_Traceback_TransIT_1989} whereas the NB-LDPC code performance continues to improve with blocklength. The gaps between the throughputs for $m=\infty$, $m=5$, and $m=10$ NB-LDPC codes are similar to the gap observed in Fig.~\ref{fig:sdartlambda} . For $m=10$ the performance of NB-LDPC codes in VLFT is much closer to the case of $m=\infty$. The NB-LDPC codes of Fig.~\ref{fig:VLFT_rtlambda} are over $GF(256)$. The shortest code for $k=96$ bits has an initial blocklength of 15 $GF(256)$ symbols (120 bits), corresponding to an initial rate of 0.8. The NB-LDPC codes for $k=192$ and $k=288$ have initial blocklengths of 256 and  384 bits, respectively. Some results for the finite-$m$ systems follow the non-active feedback scheme described in Section~\ref{sec:Creating_a_bit}.

\begin{figure}[t]
{\includegraphics[width=0.49\textwidth]
{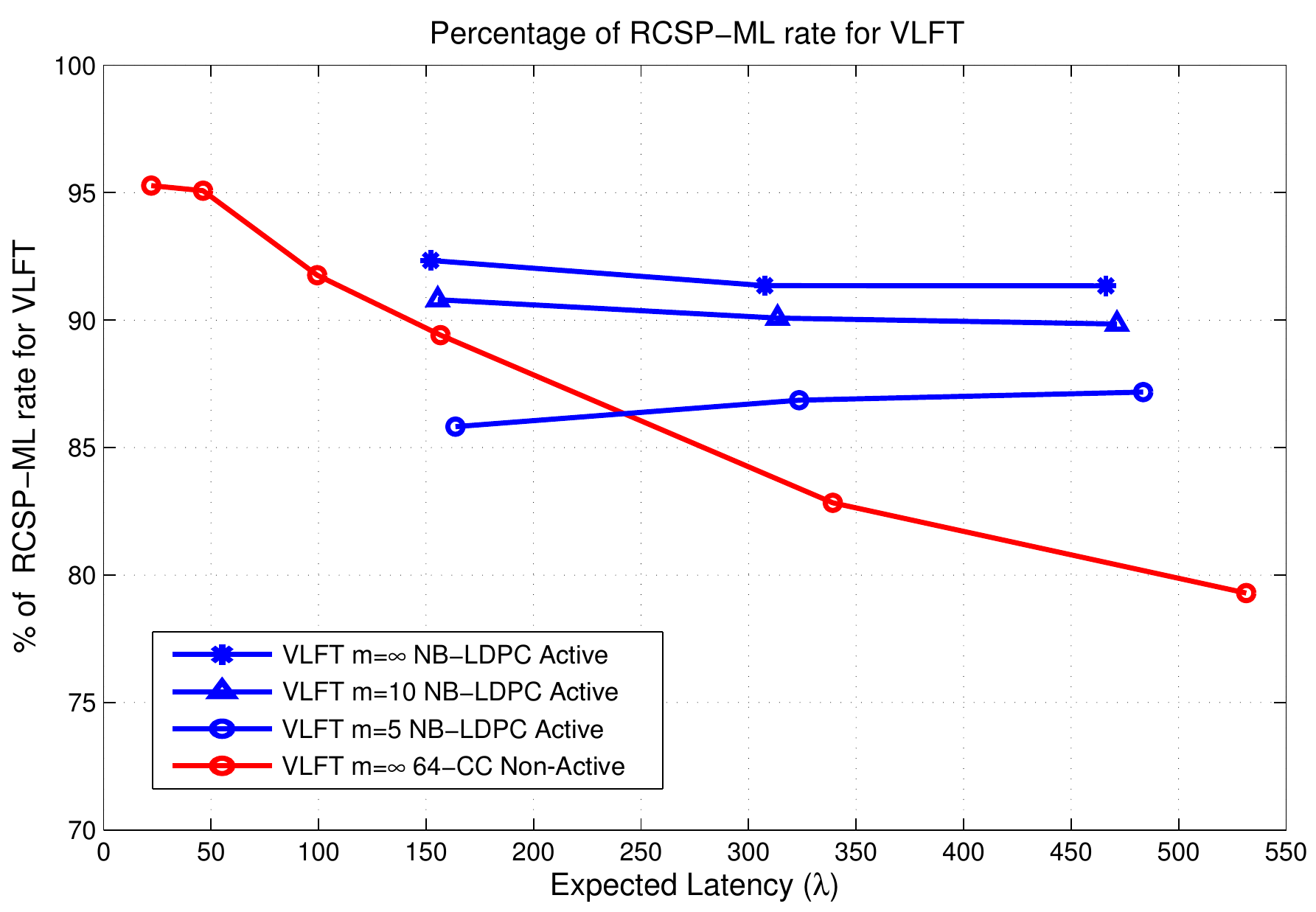}
\caption{Percentage of VLFT $R_T$ that NB-LDPC achieves with $m=\infty$, $m=10$, and $m=5$.
}\label{fig:VLFT_rtpercentage}}
\end{figure}

Fig.~\ref{fig:VLFT_rtpercentage} shows the percentage of RCSP-ML rate for VLFT achieved by NB-LDPC and convolutional codes in VLFT. In the expected-blocklength range of 150-600 bits, NB-LDPC codes achieve a throughput of about $90\%$ of RCSP-ML  throughput (and about 91\% and 96\% of unconstrained and binary-input capacity, respectively) with an unlimited number of transmissions. When the number of the transmissions is limited to $10$ and $5$, the throughput percentage decreases to about 90\% and 85\%, respectively.  RCSP-ML analysis is applied to the unconstrained-input AWGN channel at SNR 2-dB, for which the capacity is 0.684. The capacity of BI-AWGN channel at 2-dB SNR is 0.642 which is about 6\% lower than the unconstrained-input AWGN capacity.

\begin{table}[t]
 \renewcommand*{\arraystretch}{1.12}
\begin{center}
  \caption{Optimized $\{N_1,\hdots,N_5\}$ for two-phase VLF and VLF-with-CRC with $m$$=$$5$ at SNR 2 dB, and corresponding $R_T$ and $\lambda$ values achieved in simulations. $\{A_1,\hdots,A_5\}=\{5,4,3,3,3\}$ for two-phase VLF using NB-LDPC codes.   For the convolutional codes, $A_i =6$, $8$, and $9$ $\forall i$ for $k = 96$, $192$, and $288$ bits, respectively. }
   \begin{tabular}{c|c|c|c|c|c}
Code & $k$ & $\{N_1, N_2, \hdots, N_5\}$ & $\lambda$ & $R_T$ & \%\\   
\hline 
CRC NB & 89 & 143,\!  153,\!  163,\!  176,\!  201 & 164.5 &0.541 &84.2\\
2-Phase NB& 96& 146,\! 158,\! 170,\! 184,\! 211 &170.4&0.563&87.7 \\
2-Phase CC& 96&  138,\! 153,\! 166,\! 180,\!  204 & 168.6 &0.569 & 88.6 \\
\hline 
CRC NB & 185 & 293,\! 309,\! 325,\! 346,\! 386 & 323.4 &0.572 &89.1\\
2-Phase NB& 192& 301,\! 322,\! 344,\! 369,\! 408  &330.5&0.581&90.5 \\
2-Phase CC& 192& 287,\!  309,\! 331,\! 352,\! 384 &349.4&0.549&85.4\\
 \hline 
CRC NB &281& 459,\!  487,\!  518,\!  550,\!  597 & 491.3 &0.572 & 89.1\\
2-Phase NB& 288& 459,\! 487,\! 518,\! 550,\! 597  &495.7&0.581&90.5\\
2-Phase CC& 288&  416,\! 441,\! 463,\! 488,\! 532 & 599.6&0.480 &74.8 \\

\end{tabular}
\label{tbl:VLFLDPC_CC}
\end{center}
\end{table}

Table \ref{tbl:VLFLDPC_CC} summarizes the blocklengths that maximize the throughput in the two-phase VLF and VLF-with-CRC settings with $\epsilon$$=$$10^{-3}$, for both NB-LDPC codes and (for comparison) tail-biting convolutional codes. Blocklengths for the NB-LDPC codes are obtained from (\ref{eqn:VLFEK}-\ref{eqn:VLFENm}) using ES on an optimization space limited by initial SDO results.  Blocklengths for the convolutional codes are based on the coordinate-descent algorithm in \cite{Williamson2014dissertation} using the assumption of rate-compatible sphere-packing. Table \ref{tbl:VLFLDPC_CC} also shows the percentage of BI-AWGN capacity obtained in the two-phase VLF setting with $m=5$ transmissions.

 For $k=192 \text{ and } 288$, the NB-LDPC code obtains throughputs greater than $90\%$ of BI-AWGN capacity with an average blocklengths $\lambda$ of  less than $500$ bits in the 2-phase setting. NB-LDPC codes in the VLF-with-CRC setting with $m=5$ achieve throughputs slightly lower than the ones in the 2-phase setting with $m=5$. However, similar to Fig.~\ref{fig:sdartlambda} if $m$ is increased to $10$, VLF-with-CRC results in higher throughputs.  Large values of $m$ lead to a degradation in throughput performance for two-phase VLF due to the overhead associated with the more frequent forward ACK and NACK messages in the confirmation phase.

The rate-$1/3$ convolutional codes in Table \ref{tbl:VLFLDPC_CC}
have octal generator polynomials $(117, 127, 155)$ for the 64-state code and $(2325, 2731, 3747)$ for the 1024-state code  \cite{Kasraisit2014}.  The NB-LDPC codes are described completely online\footnote{UCLA Communication Systems Laboratory  (CSL) website at http://www.seas.ucla.edu/csl/resources/index.htm}.

\begin{figure}[t]
{\includegraphics[width=0.49\textwidth]{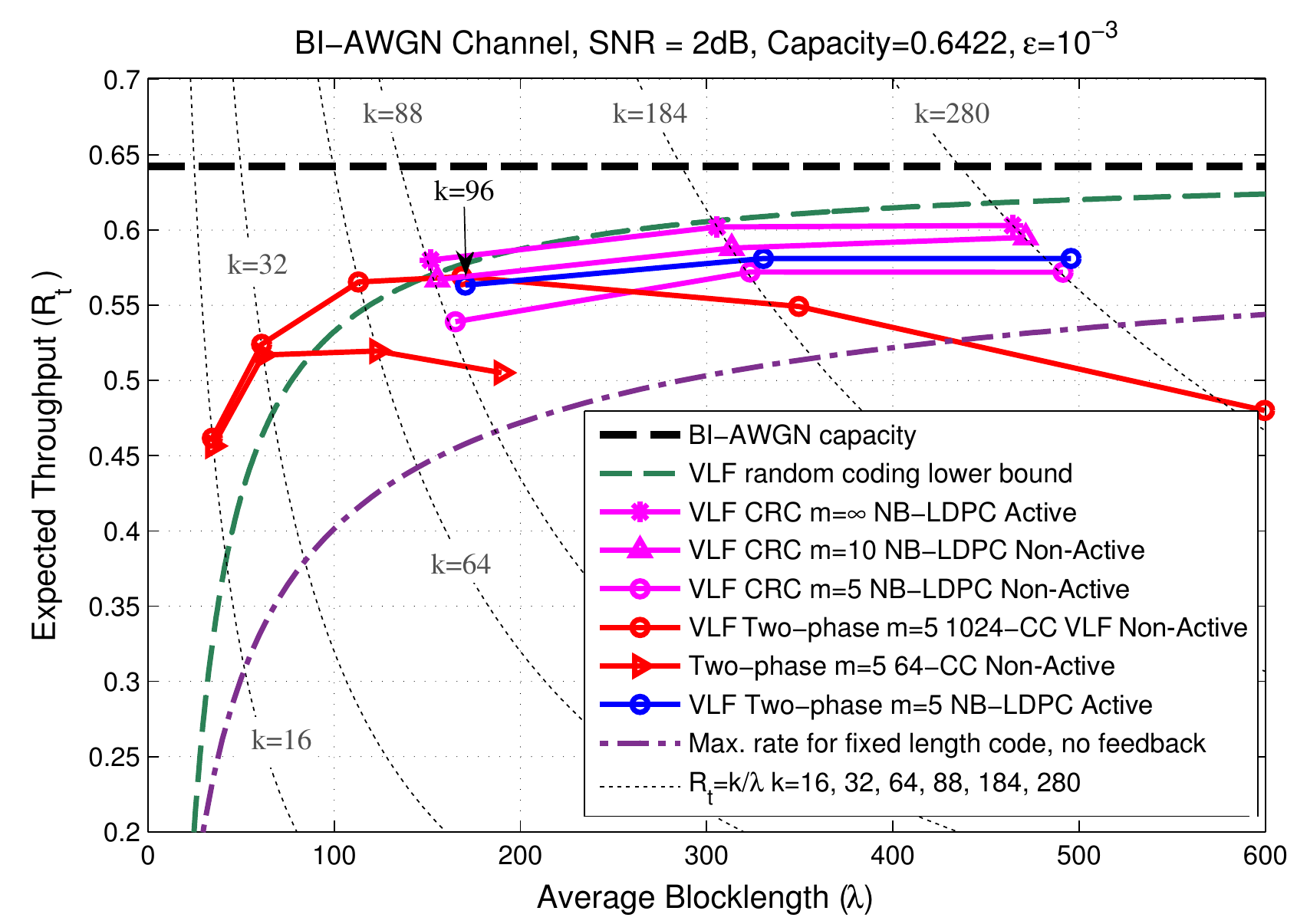}
\caption{$R_T$ vs. $\lambda$ for NB-LDPC with $m=\infty$ in VLF-with-CRC and 64 and 1024-state convolutional codes and NB-LDPC codes with $m=5$ in VLF.
}\label{fig:VLF_rtlambda}}
\end{figure}

Fig.~\ref{fig:VLF_rtlambda} shows the throughput obtained in the VLF setting for NB-LDPC codes, 64-state and 1024-state tail-biting convolutional codes with $m=5$, $m=10$, $m=\infty$ for $\epsilon=10^{-3}$. As the blocklength increases, as mentioned in \cite{Chen_Feedback_Journal_2013}, the performance of the codes in VLF gets closer to the performance in VLFT. The plots for $m=5$ are from Table \ref{tbl:VLFLDPC_CC}. With $m=\infty$, the $k=89$ the NB-LDPC code achieves a throughput greater than the random coding lower bound obtained from the analysis in \cite{Polyanskiy_IT_2011_NonAsym}.


\begin{figure}[t]
{\includegraphics[width=0.49\textwidth]
{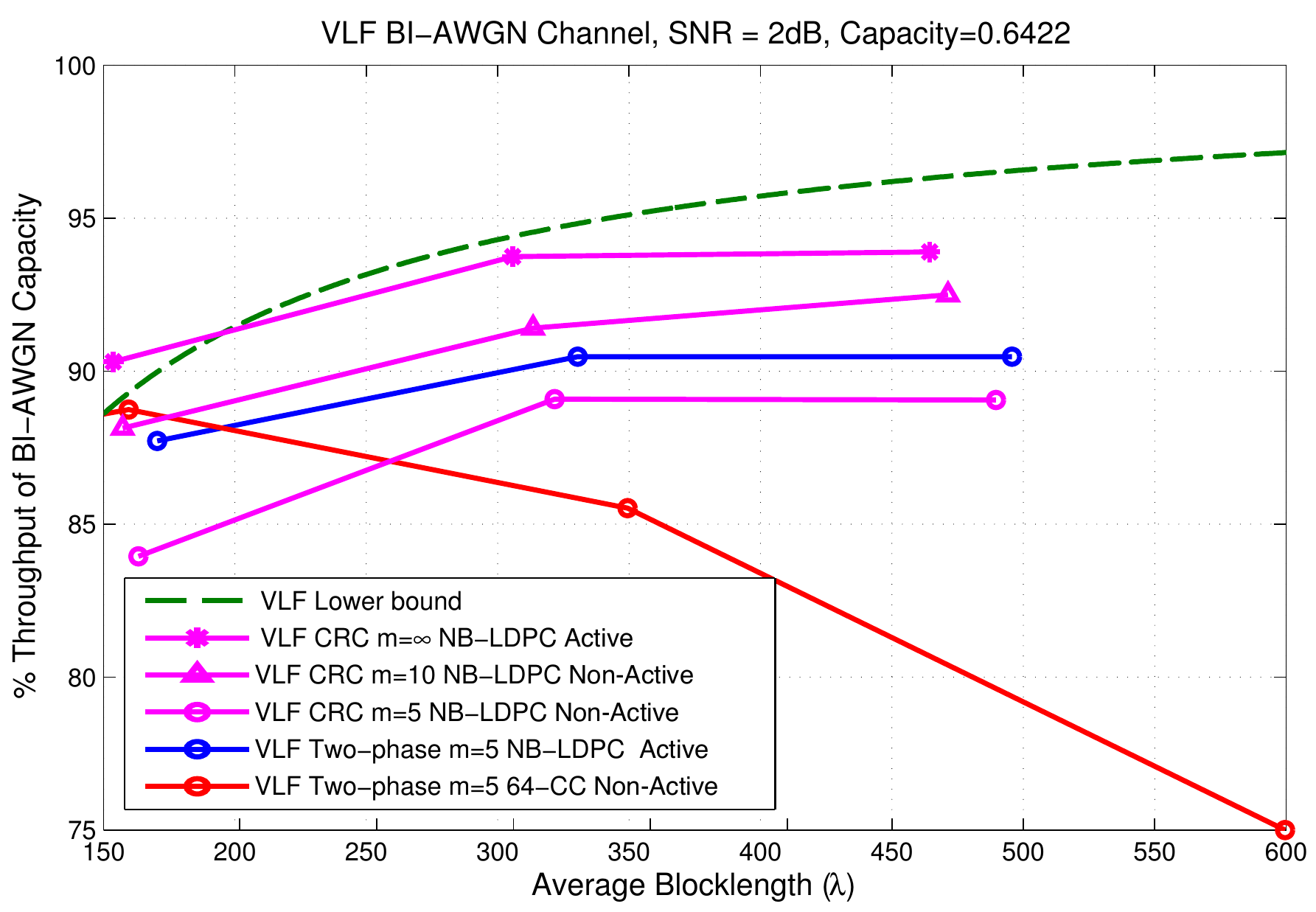}
\caption{Percentage of BI-AWGN capacity that  NB-LDPC and convolutional codes achieve in VLF. 
}\label{fig:VLF_rtpercentage}}
\end{figure}

Fig.~\ref{fig:VLF_rtpercentage} shows the percentage of the capacity of the BI-AWGN channel at 2-dB SNR achieved by NB-LDPC and convolutional codes using VLF.  In the expected blocklength range of 300-500 bits, NB-LDPC codes with CRC achieve a throughput of about $94\%$ of capacity with an unlimited number of transmissions. When the number of the transmissions is limited to $10$, the throughput percentage decreases to about 93\%. For $m=5$, NB-LDPC codes perform slightly better in two-phase VLF setting than in VLF-with-CRC.  Note that for $m=\infty$ or even $m=10$ two-phase VLF will not perform well because of the overhead associated with the confirmation messages.  

As discussed in Section~\ref{sec:incremental} similar Gaussian approximation analysis can be done for higher-SNR AWGN channels. for instance, for SNR-8dB AWGN channel which uses a larger 16-QAM constellation, the VLF-with-CRC system with an unlimited number of transmissions achieves a throughput of 2.37 bits per symbol with a frame error probability of less than $10^{-3}$. This throughput corresponds to 88\% of capacity in the blocklength regime of 40 16-QAM (quadrature amplitude modulation) symbols. Furthermore, the VLF-with-CRC system on 5-dB BI-AWGN fading channel with an unlimited number of transmissions achieves a throughput corresponding to 90\% of capacity in the blocklength regime of about 140 bits.


\section{conclusion}
\label{sec:conclusion}
This paper uses the reciprocal-Gaussian approximation for the blocklength of first successful decoding to optimize the size of each incremental transmission to maximize throughput in VLFT and VLF settings. For feedback with a limitation on the number of transmissions, the sequential differential optimization (SDO) algorithm can be used quickly and accurately to find the optimal transmission lengths for a wide range of channels and codes.  In this paper we applied SDO to non-binary LDPC codes for a variety of feedback systems.  We focused on the binary-input AWGN channel but verified the effectiveness of the Gaussian approximation and SDO on the standard AWGN channel with a 16-QAM input and on a fading channel. In the 300-500 bit average blocklength regime, this paper reports the best VLFT and VLF throughputs yet.  VLFT throughputs are higher than VLF, but VLF is more practical because it does not assume a noiseless transmitter confirmation symbol. For VLF-with-CRC with $m=\infty$, NB-LDPC codes with optimized blocklengths achieve about $94\%$ of the capacity of 2-dB BI-AWGN channel for an average blocklength of 300-500 bits. In the same blocklength regime, for VLF-with-CRC with $m=10$, NB-LDPC codes with optimized blocklengths achieve about $93\%$ of the capacity.   

The performance results can also be considered in terms of SNR gap.  In Fig.~\ref{fig:VLF_rtpercentage}, the random-coding lower bound for a system with feedback is 0.27 dB from the Shannon limit  for $k=280$ with a blocklength of less than 500 bits.  Looking at the VLF-CRC NB-LDPC codes for $k=280$ in Fig.~\ref{fig:VLF_rtpercentage}, the $m=\infty$ NB-LDPC code is 0.53 dB from Shannon limit.  The NB-LDPC non-active feedback system in Fig.~\ref{fig:VLF_rtpercentage} uses ten rounds of single-bit feedback to operate within 0.65 dB of the Shannon limit with an average blocklength of less than 500 bits.
Similar analysis can also be done for higher-SNR AWGN and fading channels.

%
%
\bibliographystyle{IEEEtranTCOM}
{\bibliography{KV_bib}}
\end{document}